\documentclass[fleqn,usenatbib]{mnras}

\usepackage{newtxtext,newtxmath}

\usepackage[T1]{fontenc}
\usepackage{ae,aecompl}

\usepackage{graphicx}	
\usepackage{amsmath}	
\usepackage{amssymb}	

\newcommand{\kmspc}{\,km\,s$^{-1}$\,pc$^{-1}$}	
\defcitealias{2010ApJ...711.1017P}{P10}

\title[Ionised gas kinematics in bipolar H\,{\normalsize \textit{II}} regions]
{Ionised gas kinematics in bipolar H\,{\Large \textbf{II}} regions}

\author[Hannah Dalgleish et al.]{Hannah S. Dalgleish$^{1}$\thanks{E-mail: h.s.dalgleish@2016.ljmu.ac.uk},
Steven N. Longmore$^{1}$,
Thomas Peters$^{2}$, \newauthor
Jonathan D. Henshaw$^{1,3}$, 
Joshua L. Veitch-Michaelis$^{4}$, and   
James S. Urquhart$^{5}$
\\
$^{1}$Astrophysics Research Institute, Liverpool John Moores University, Liverpool, L3 5RF, UK\\
$^{2}$Max-Planck-Institut f{\"u}r Astrophysik, Karl-Schwarzschild-Str. 1, 85748 Garching, Germany\\
$^{3}$Max-Planck-Institut f{\"u}r Astronomie, K{\"o}nigstuhl 17, 69117 Heidelberg, Germany \\
$^{4}$I3D Robotics Ltd, Tonbridge, TN9 1SP, UK \\
$^{5}$Centre for Astrophysics and Planetary Science, University of Kent, Canterbury, CT2 7NH, UK
}
 
\date{Accepted XXX. Received YYY; in original form ZZZ}

\pubyear{2018}

\begin{document}
\label{firstpage}
\pagerange{\pageref{firstpage}--\pageref{lastpage}}
\maketitle

\begin{abstract}
Stellar feedback plays a fundamental role in shaping the evolution of galaxies. Here we explore the use of ionised gas kinematics in young, bipolar \ion{H}{II} regions as a probe of early feedback in these star-forming environments. We have undertaken a multi-wavelength study of a young, bipolar \ion{H}{II} region in the Galactic disc, G316.81--0.06, which lies at the centre of a massive ($\sim10^3$ M$_{\sun}$) infrared-dark cloud filament. It is still accreting molecular gas as well as driving a $\sim0.2$ pc ionised gas outflow perpendicular to the filament. Intriguingly, we observe a large velocity gradient ($47.81 \pm 3.21$ \kmspc) across the ionised gas in a direction perpendicular to the outflow. This kinematic signature of the ionised gas shows a reasonable correspondence with the simulations of young \ion{H}{II} regions. Based on a qualitative comparison between our observations and these simulations, we put forward a possible explanation for the velocity gradients observed in G316.81--0.06. If the velocity gradient perpendicular to the outflow is caused by rotation of the ionised gas, then we infer that this rotation is a direct result of the initial net angular momentum in the natal molecular cloud. If this explanation is correct, this kinematic signature should be common in other young (bipolar) \ion{H}{II} regions. We suggest that further quantitative analysis of the ionised gas kinematics of young \ion{H}{II} regions, combined with additional simulations, should improve our understanding of feedback at these early stages. 
\end{abstract}

\begin{keywords}
ISM: \ion{H}{II} regions; kinematics and dynamics - stars: massive; protostars
\end{keywords}



\section{Introduction}
\label{intro}

Feedback from high-mass stars (i.e. OB stars with M$_{\star}$ $\geq$ 8 M$_{\sun}$) is fundamental to the shaping of the visible Universe. From the moment star formation begins, stellar feedback commences, injecting energy and momentum into the natal environment. This feedback can both hinder and facilitate star formation; negative feedback restrains or can even terminate star formation, whereas positive feedback acts to increase the star formation rate and/or efficiency. Many different physical mechanisms contribute to feedback by varying degrees, each depending on a variety of factors (e.g. initial conditions), resulting in an intricate and interdependent series of processes. In a recent review on this topic, \citet{2014prpl.conf..243K} groups feedback processes into three main categories: momentum feedback (e.g. protostellar outflows and radiation pressure); ``explosive'' feedback (e.g. stellar winds, photoionising radiation, and supernovae); and thermal feedback (e.g. non-ionising radiation).

Stellar feedback encompasses many astrophysical processes, moderating star formation from stellar scales ($\ll$ 1 pc) to cosmological kpc-scales (e.g. driving Galactic outflows; \citealt{2011ApJ...735...66M,2016MNRAS.456.3432G}). 
Despite our growing knowledge of these processes, the overarching interplay between them remains uncertain. Observationally, limited spatial resolution makes it difficult to disentangle the effects of feedback mechanisms which operate simultaneously. Other additional factors, such as the role of magnetic fields (for which the strength and orientation are difficult to measure) and feedback from surrounding low-mass stars, complicate the process further. Moreover, limited observations of the earliest stages of high-mass star formation means that the large samples needed for a robust statistical analysis are lacking. With observatories like ALMA and the EVLA, which have sufficient angular resolution to resolve and detect individually forming high-mass stars, our understanding is continually improving.

Meanwhile, in the past few decades, there have been considerable efforts attempting to simulate the vast range of stellar feedback effects. It has been clearly demonstrated that without feedback, simulations fail to replicate the galaxies that we observe in the Universe today (e.g. \citealt{1996ApJS..105...19K,1999MNRAS.310.1087S,2000MNRAS.319..168C,2003MNRAS.339..312S,2009MNRAS.396.2332K,2011MNRAS.413.2741G,2012ARA&A..50..531K,2012MNRAS.423.1726S,2014MNRAS.445..581H,2017MNRAS.466.3293P}) and often produce galaxies that are much more massive than observed.
Consequently, simulations have looked to feedback for answers, with promising results. Yet it remains a great challenge to create a model which includes all feedback processes over a vast range of scales. Often only one or two types of feedback are included (e.g. \citealt{2010ApJ...713.1120K,2011ApJ...735...49M,2013MNRAS.430..234D,2013ApJ...770...25A,2013ApJ...776....1K,2014ApJ...788...14P,2015ApJ...801...33T,2015MNRAS.454.2691M,2016ApJ...824...79A,2017ApJ...841...82B,2017ApJ...836..204N}).
In order to improve simulations and implement more feedback effects, better understanding of the relevant physical processes is needed. This will help to provide the observational constraints needed for parameterising the simulations.

\subsection
[{H II regions}]
{H\,{\sevensize II} regions}
The study of \ion{H}{II} regions can allow us to explore how high-mass stars impact their environment via the aforementioned feedback mechanisms. \ion{H}{II} regions are bright in the radio regime, particularly with radio recombination lines (RRLs) and thermal bremsstrahlung, both clear diagnostics of high-mass star formation. See \citet{2017arXiv171105275H} for a review on synthetic observation studies, particularly regarding feedback and the global structure of \ion{H}{II} regions.
 
Predominantly, the study of \ion{H}{II} regions has focused on surveys examining morphologies, sizes and densities (e.g. \citealt{2006AJ....131.2525H,2012PASP..124..939H,2013MNRAS.431.1752U,2013MNRAS.435..400U,2017A&A...602A..37K,2017A&A...603A..33G}). 
In terms of morphology, ultracompact (UC; $\leq 0.1$ pc) \ion{H}{II} regions can be categorised as either spherical, cometary, core-halo, shell, or irregular \citep{1989ApJS...69..831W}. \citet{2005ApJ...624L.101D} modified the classification scheme to also include bipolar morphologies. 
The work of \citet{1989ApJS...69..831W} found that too many \ion{UCH}{II} regions are observed considering their short apparent lifetime, known as the `lifetime debate'.
\citet{2010ApJ...719..831P} propose a solution based on their synthetic radio continuum observations of young high-mass star formation regions. They found that \ion{H}{II} regions `flicker' as they grow, a result of a fluctuating accretion flow around the high-mass star (fragmentation-induced starvation; \citealt{2010ApJ...711.1017P}, hereafter \citetalias{2010ApJ...711.1017P}). This is a possible resolution to the lifetime problem since the young \ion{H}{II} regions shrink and grow rapidly as they evolve. Short (several year) variations in the flux density of the high-mass star forming region Sgr B2 have been observed \citep{2014ApJ...781L..36D,2015ApJ...815..123D}, which the authors attribute to this `flickering'.

There have also been detailed studies on the kinematics of \ion{H}{II} regions on (proto)stellar scales ($\lesssim$ 10,000 AU). This includes the accretion of ionised material onto forming high-mass stars (e.g. \citealt{1988ApJ...324..920K,2008ApJ...678L.109K,2005ApJ...624L..49S,2006ApJ...637..850K,2002ApJ...568..754K,2003ApJ...599.1196K,2007ApJ...666..976K,2008ApJ...674L..33G,2017arXiv171204735K}); the gravitational collapse and rotation of turbulent molecular clouds (e.g. \citealt{2000ApJ...535..887K,2009ApJ...703.1308K}); ionised outflows (e.g. \citealt{1994ApJ...428..670D,2013A&A...556A.107K,2016ApJ...818...52T}); and the rotation of ionised gas on stellar scales (e.g. \citealt{1994ApJ...428..324R,2008ApJ...681..350S}).

However, to date, fewer studies have been devoted to measuring the ionised gas kinematics of \ion{H}{II} regions on cloud scales ($\sim$ 0.1 pc), pertinent to understanding the effect of feedback of high-mass stars on their natal clouds. Most studies have focused on understanding the kinematics of cometary \ion{H}{II} regions in order to deduce which model (e.g. bow shock or champagne) applies. This is often done via the analysis of velocity ranges or gradients of the ionised gas (e.g. \citealt{1996ApJ...464..272L,1999MNRAS.305..701L,2017MNRAS.465.4219V}).
Intriguingly, G34.3+0.2, G45.07+0.13 and Sgr B2 I and H all show ranges in velocity (from 10-35 km s$^{-1}$), perpendicular to the axis of symmetry of the cometary \ion{H}{II} region \citep{1986ApJ...309..553G,1990ApJ...351..538G,1994ApJ...432..648G,2014A&A...563A..39I}.

Kinematic studies of \ion{H}{II} regions across all morphological types could present new interpretations on our understanding of feedback. For example, bipolar \ion{H}{II} regions could provide an insight to early feedback, since they typically exist at earlier evolutionary stages when ionisation has only just begun (e.g. \citealt{2010ApJ...721..222B}). Newly ionised material flows outwards with velocities up to 30 km s$^{-1}$ \citep{2015A&A...582A...1D} and neutral material (usually in the form of a molecular disc) lies perpendicular to the outflows, often showing signs of accretion towards a central (proto)star. When viewed approximately edge-on, the \ion{H}{II} region appears as bipolar. Velocity gradients within the ionised gas will typically correspond to infall, outflow, rotation or a combination, which will be influenced by the viewing angle. This can have different implications for feedback, depending on which motion truly occurs.

In this paper, we present observations of a young, bipolar \ion{H}{II} region, G316.81--0.06. Our results show a velocity gradient in the ionised gas at 0.1 pc scales, perpendicular to the bipolar axis. In conjunction with the \citetalias{2010ApJ...711.1017P} simulations, we aim to understand the origin of the velocity structure in the ionised gas and its relation to feedback.
Section \ref{data} describes the observations and simulations in more detail, followed by the data analysis in \S\, \ref{method}. The results and discussion are in \S\S\, \ref{results} and \ref{dis}, concluding with a summary in \S\, \ref{summary}.


\begin{figure*}
	\centering
    	\includegraphics[width=0.99\textwidth]{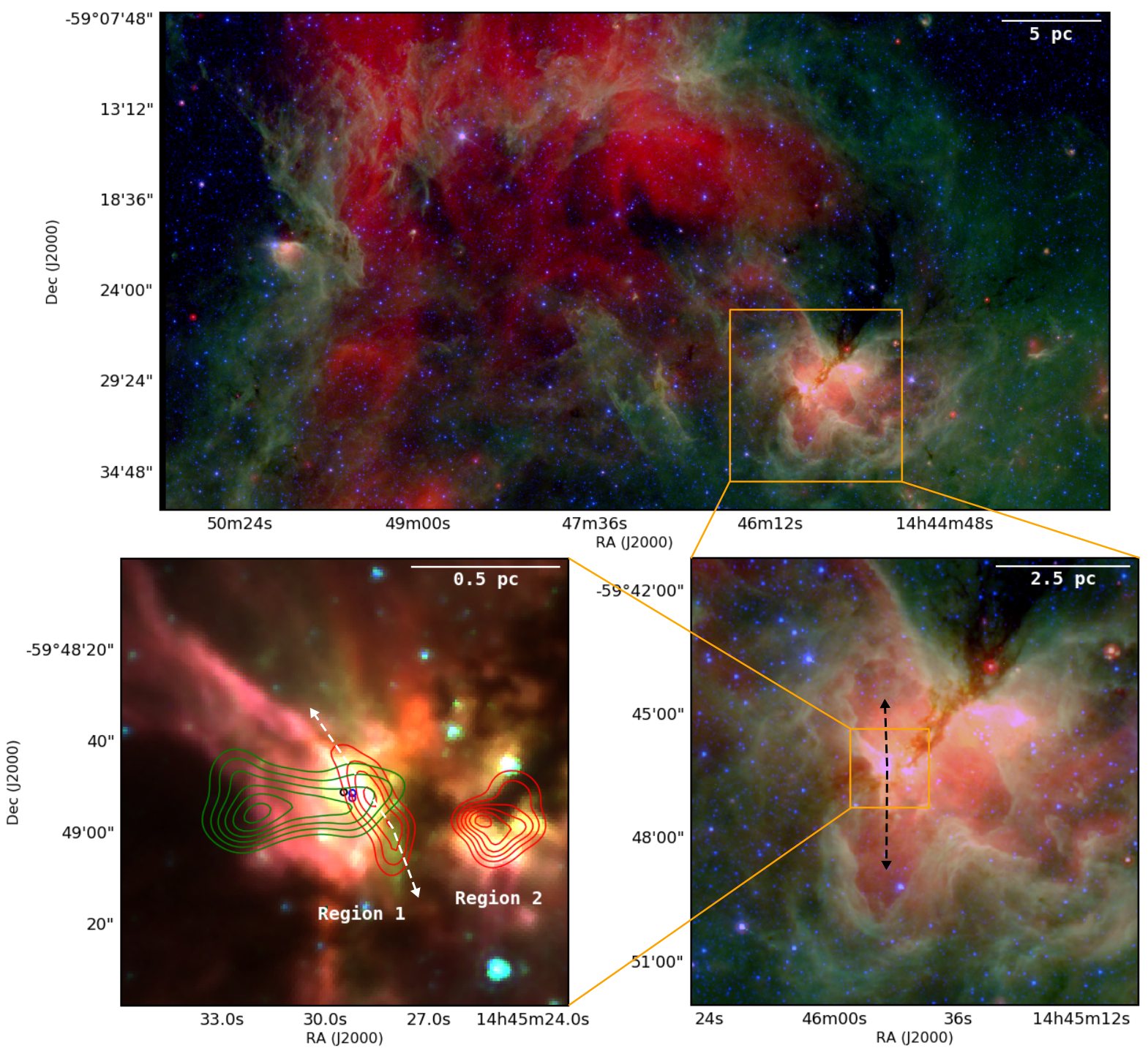}
	\caption{Multi-wavelength images of G316.81--0.06. Top and bottom-right: \textit{Spitzer} GLIMPSE/MIPSGAL image in 3.6, 8.0, and 24.0 micron IRAC bands \citep{2003PASP..115..953B,2009PASP..121..213C,2009PASP..121...76C,2015AJ....149...64G,2012ascl.soft06002L}. The dashed black arrows indicate an outflow in the direction of the two north-south MIR bubbles.
Bottom-left: GLIMPSE image in 3.6, 4.5, and 8.0 micron IRAC bands \citep{2003PASP..115..953B,2009PASP..121..213C}. Red and green contours show the 35-GHz continuum \citep{2009MNRAS.399..861L} and integrated NH$_3$(1,1) \citep{1997MNRAS.291..261W} respectively. Two separate \ion{H}{II} regions are labelled Region 1 and 2 accordingly. Masers from \citet{2010MNRAS.406.1487B} are depicted as circles: blue (hydroxyl); black (water); purple (methanol). Dashed white arrows indicate the direction of an ionised outflow, aligned with the 35-GHz continuum of Region 1 and the ``green fuzzy''.}
    \label{fig:image}
\end{figure*}

\section{Data}
\label{data}
\subsection{Observations}
\label{sec:obs}

Figure \ref{fig:image} illustrates multi-wavelength images of G316.81--0.06, located 2.6 kpc away in the Galactic Disc (\citealt{2011MNRAS.417.2500G}; note that this is a newer distance estimate as opposed to the measurement of 2.7 kpc used in previous literature). Various authors have discussed the kinematic distance ambiguity in relation to this source \citep{1981A&A...102..225S,2006MNRAS.366.1096B,2014A&A...569A.125H}, and conclude it is at the near kinematic distance.

The top infrared (IR) image of Figure \ref{fig:image} 
is a \textit{Spitzer} GLIMPSE/MIPSGAL image in the 3.6, 8.0, and 24.0 micron IRAC bands \citep{2003PASP..115..953B,2009PASP..121..213C,2009PASP..121...76C,2015AJ....149...64G,2012ascl.soft06002L}. On the bottom-right a close-up of G316.81--0.06 is shown, of the same GLIMPSE/MIPSGAL image.
The region is enlarged further in the bottom-left; a mid-infrared (MIR; 3.6, 4.5, and 8.0 microns) GLIMPSE image.

On large scales, strong absorption is featured roughly SE-NW in both IR images, i.e. infrared dark clouds (IRDCs; e.g. \citealt{1998ApJ...494L.199E}). 
Emission features (MIR bright bubbles) are seen to the north and south, with one distinct and bright MIR central source found at the apex of these two bubbles. 
Using the \textit{Australian Telescope Compact Array (ATCA)}, two radio continuum sources classified as \ion{UCH}{ii} regions in \citet{1997MNRAS.291..261W,1998MNRAS.301..640W} are overlaid with red contours (bottom-left image) showing the 35-GHz continuum \citep{2009MNRAS.399..861L}. The left-hand source (Region 1), shows two distinct lobes elongated roughly NE-SW. The continuum data were taken in addition to the H70$\upalpha$ RRL with a compact antenna configuration (12.5$\arcsec$ angular resolution) and thus, spatial filtering is not a major issue (see \citealt{2009MNRAS.399..861L} for further details). 

Region 1 has many more significant features. Numerous masers (hydroxyl, class II methanol, and water) have been detected, see Appendix \ref{tab:masers} for a complete list. For clarity, only the masers listed by \citet{2010MNRAS.406.1487B} are marked (bottom-left of Figure \ref{fig:image}): blue, purple, and black circles are hydroxyl, class II methanol, and water masers respectively.  
Ammonia emission (green contours; \citealt{1997MNRAS.291..261W}) coincides with the three masers, and NH$_3$ (1,1) shows a clear inverse P-Cygni profile towards the cm-continuum source which extends eastwards from Region 1 and peaks towards the IRDC \citep{2007MNRAS.379..535L}. 
Other features include 4.5 $\umu$m excess emission, i.e. a ``green fuzzy" (otherwise known as an extended green object, EGO; \citealt{2008AJ....136.2391C}) in the MIR (bottom-left of Figure \ref{fig:image}; \citealt{2007prpl.conf..165B,2009ApJS..184..366B}).

Overall, G316.81--0.06 is a very complex region, affected by contributions from multiple feedback mechanisms.
We interpret the aforementioned features as follows: 
(a) Two MIR bright sources are two separate \ion{H}{II} regions (Regions 1 and 2) -- formed from the IRDC filament -- which drive the MIR bubbles. It appears as though these cavities have been driven by an older outflow (indicated by the black dashed arrows; bottom-right of Figure \ref{fig:image}) in a north-south direction, perpendicular to the elongated ammonia emission. 
(b) Masers indicate youth (class II 6.7-GHz maser emission suggests a possible age of 10-45 kyr; \citealt{2010MNRAS.401.2219B}). 
(c) The inverse P-Cygni profile implies infall towards Region 1. 
(d) We infer a more recent outflow from the presence of the ``green fuzzy" \citep{2009ApJS..181..360C}. In combination with the elongated 35-GHz continuum, the outflow appears to be bipolar, possibly in the form of an ionised jet (indicated by the white dashed arrows; bottom-left of Figure \ref{fig:image}). 

In contrast, Region 2 lacks masers and ammonia emission. This implies that Region 2 is older than Region 1, as concluded by \citet{2007MNRAS.379..535L}. 

\begin{figure*}
	\centering
	\includegraphics[width=0.495\textwidth]{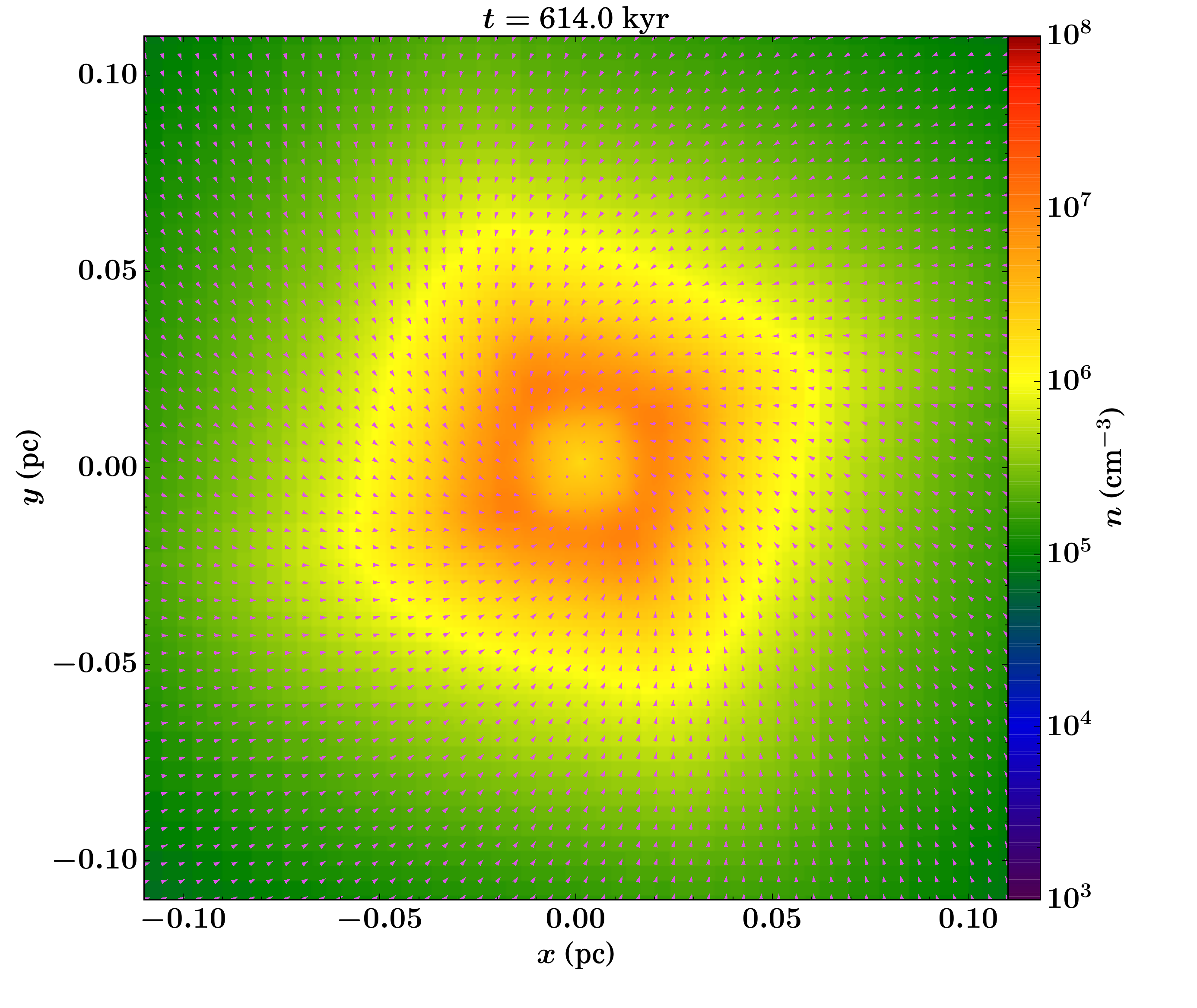}
    \includegraphics[width=0.495\textwidth]{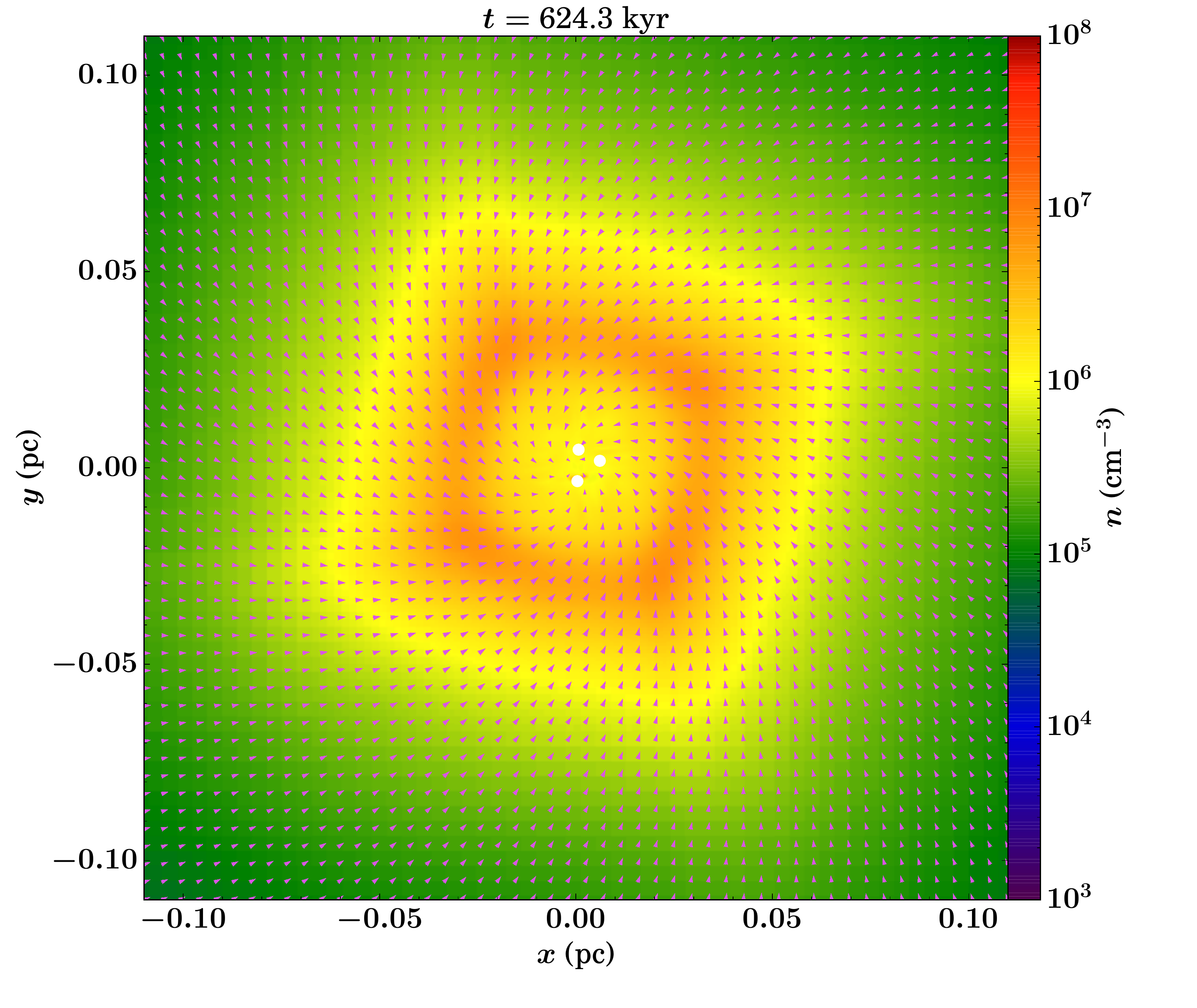}
	\includegraphics[width=0.495\textwidth]{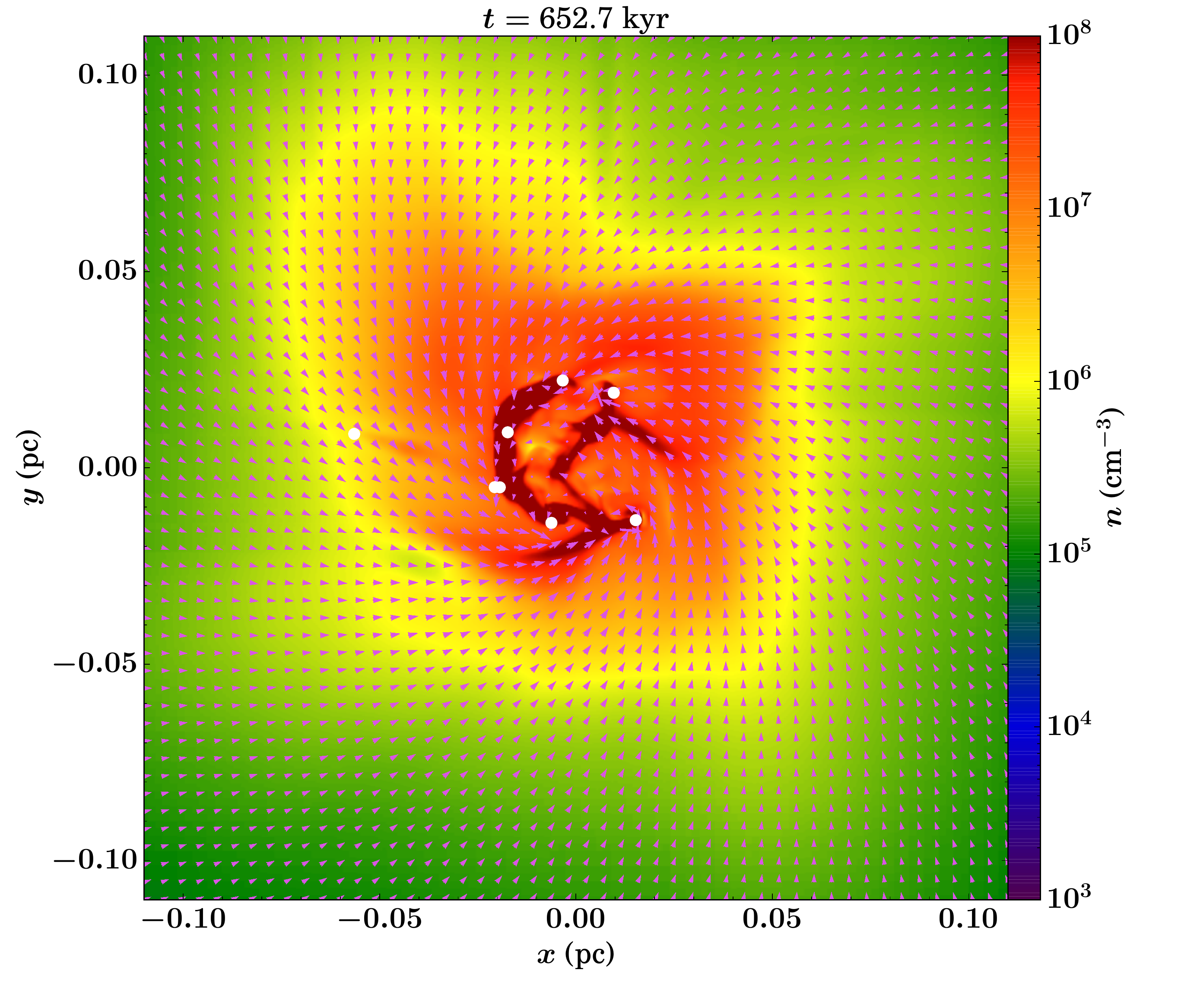}
    \includegraphics[width=0.495\textwidth]{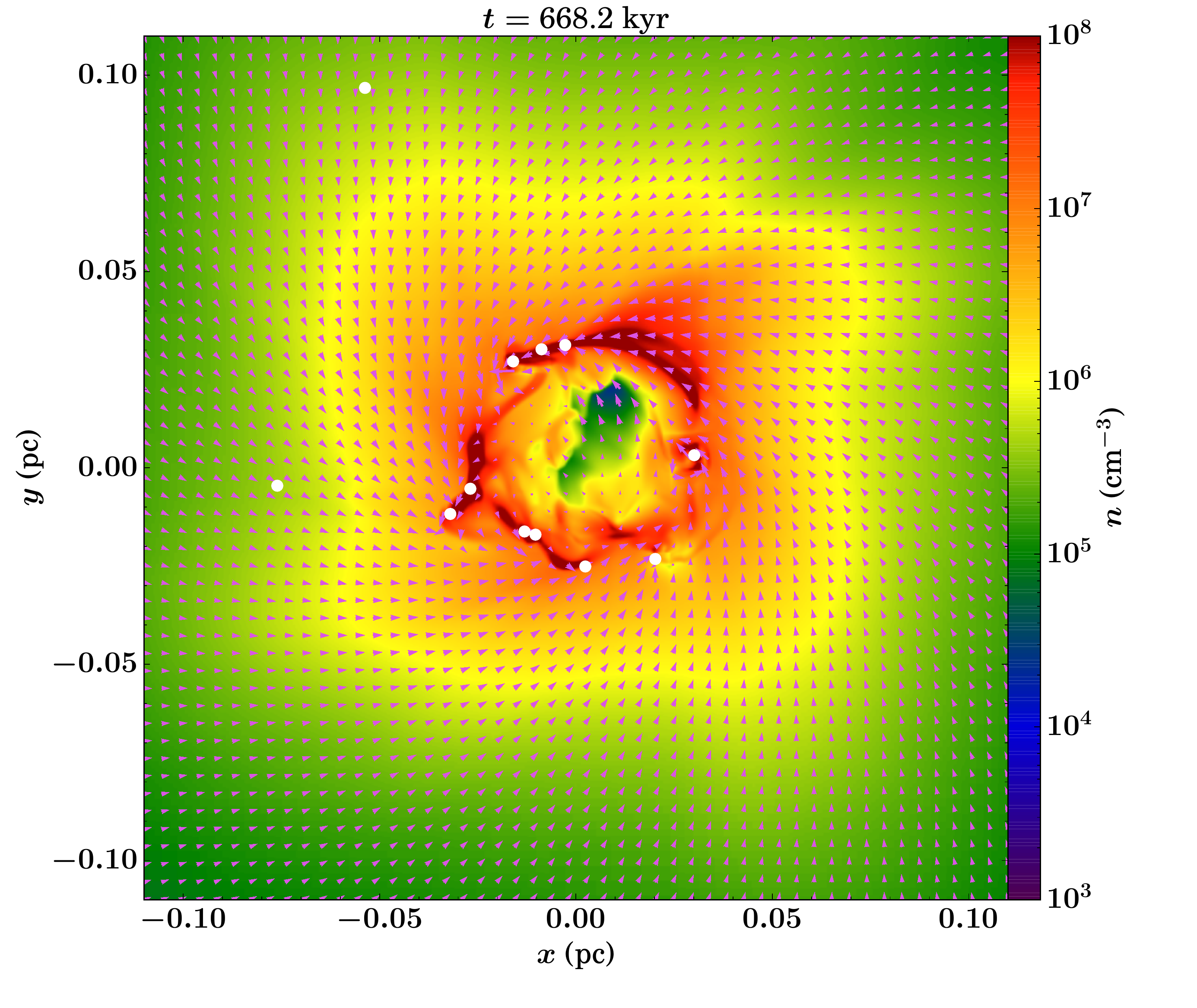}
    \caption{Snapshot density slices through the simulations of \citetalias{2010ApJ...711.1017P} depicting the stages prior to the formation of an \ion{H}{II} region in the xy-plane. The time-steps shown reflect four initial evolutionary stages of the simulation occurring at 614.0, 624.3, 652.7, and 668.2 kyr. The arrows are velocity vectors and the white points are sink particles.}
    \label{bubbles}
\end{figure*}

\begin{figure}
	\centering
	\includegraphics[width=0.495\textwidth]{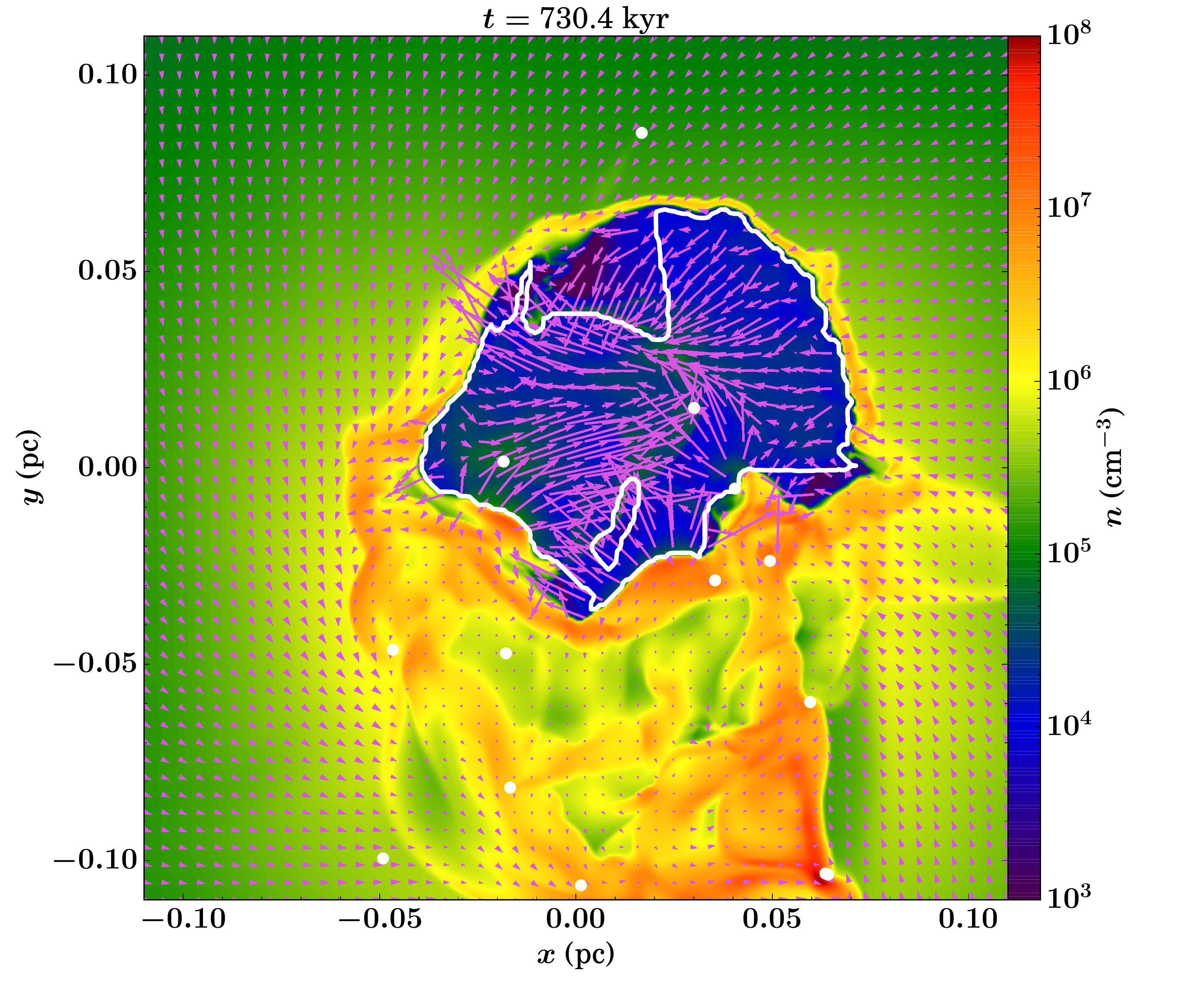}
    \includegraphics[width=0.495\textwidth]{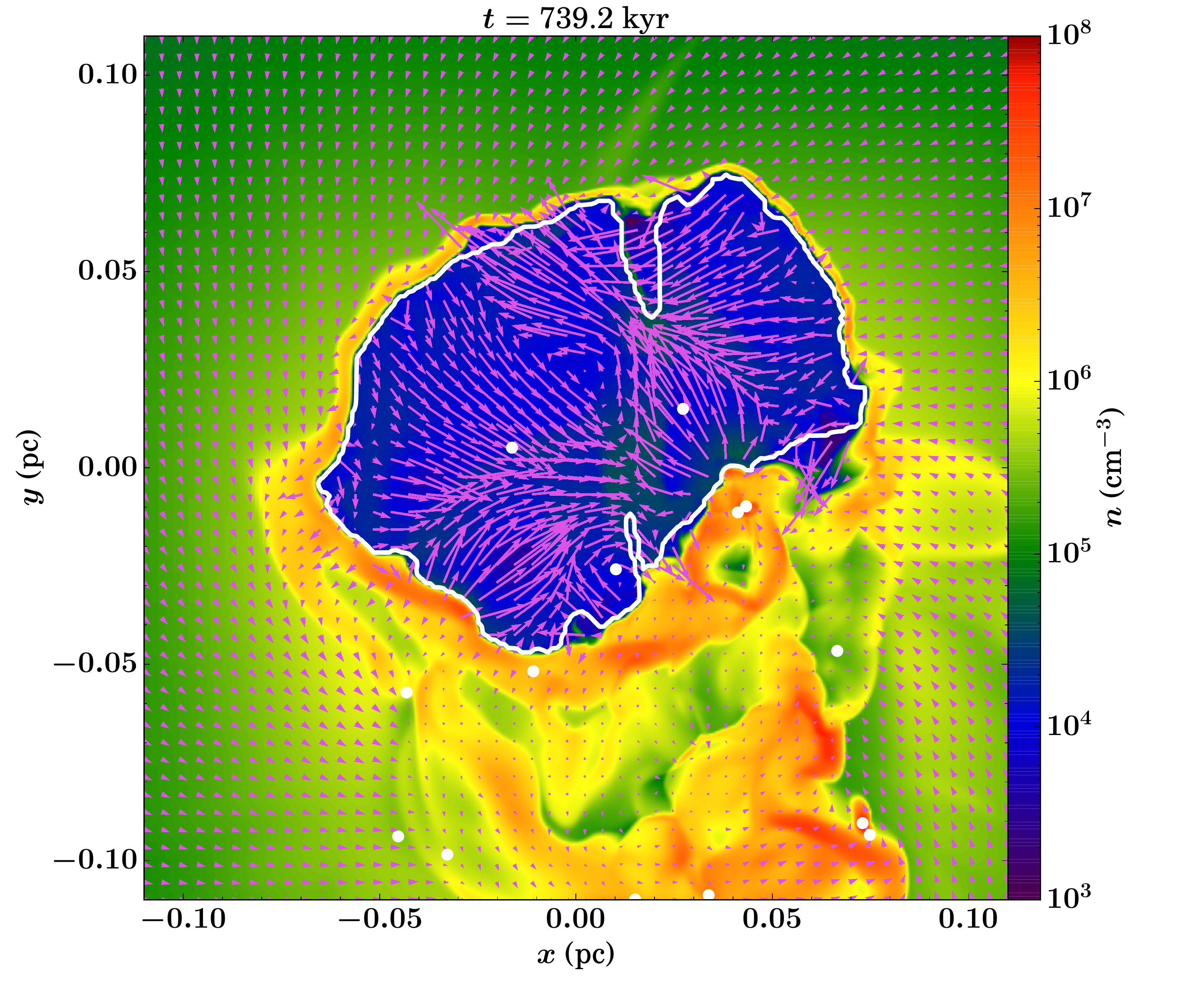}
	\includegraphics[width=0.495\textwidth]{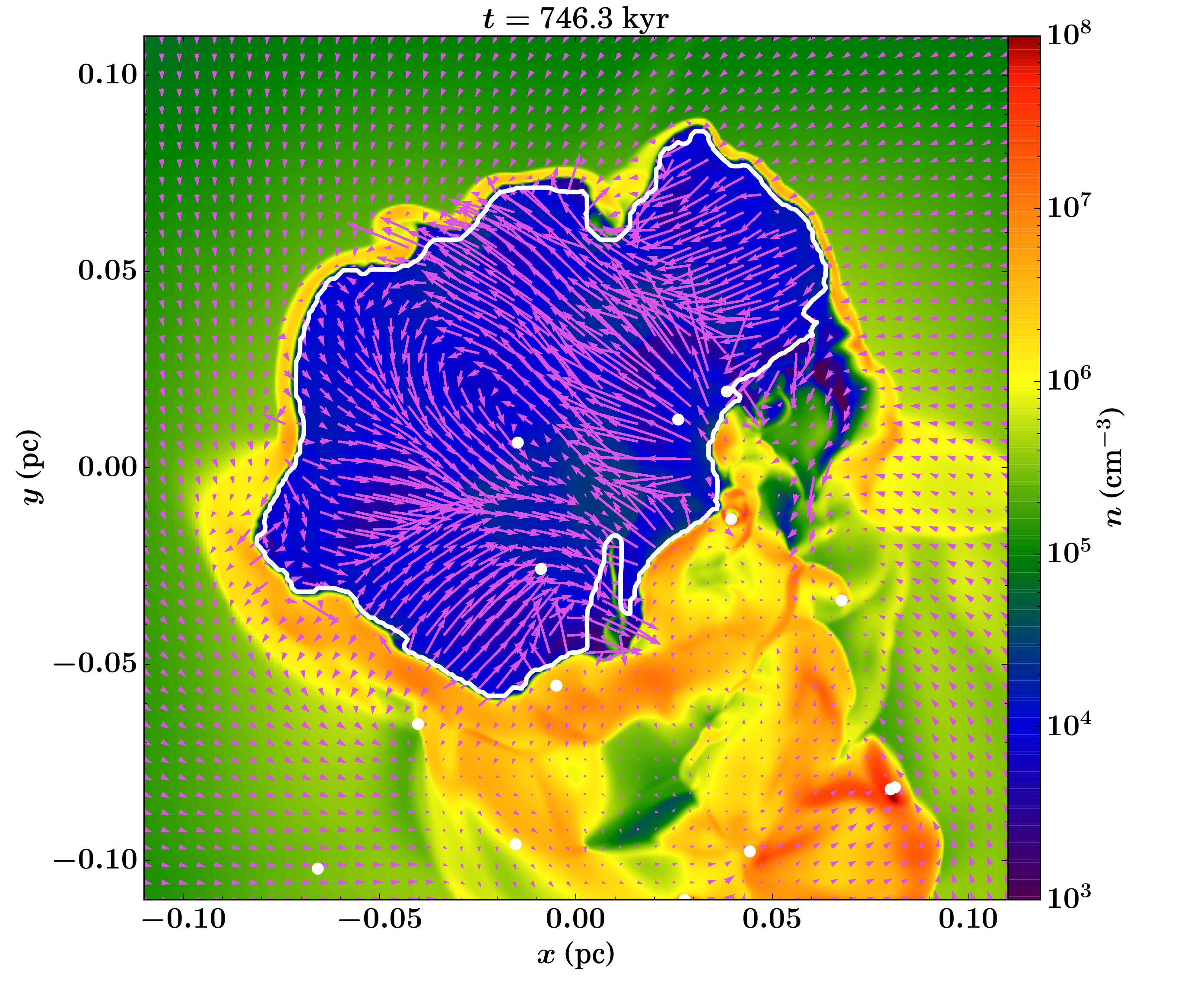}
	\caption{Snapshot density slices through the simulations of \citetalias{2010ApJ...711.1017P} after the formation of an \ion{H}{II} region in the xy-plane. The time-steps reflect later evolutionary stages which occur at 730.4, 739.2, and 746.3 kyr. The arrows depict velocity vectors and the white points are sink particles. The thin white border marks the boundary of 90\% ionisation fraction.}
    \label{finalbubbles}
\end{figure}

\subsection{Numerical Simulations}
\label{sims}
To help interpret our data we looked for numerical simulations of young \ion{H}{II} regions which match Region 1 as closely as possible. As described below, we identified the simulations of \citetalias{2010ApJ...711.1017P} as having similar global properties to G316.81--0.06, so we focus on comparing our data to these simulations. The \citetalias{2010ApJ...711.1017P} simulations have not been fine-tuned to the observations. 

Given our limited knowledge of the G316.81--0.06 region's history, and with only a single observational snapshot of the region's evolution, it is impossible to know how closely the initial conditions of the simulations are matched to the progenitor gas cloud of G316.81--0.06. For that reason our approach is to try and identify general trends in the evolution of the simulations in the hope that the underlying physical mechanisms driving this evolution will be applicable to the largest number of real \ion{H}{II} regions. We avoid focusing on detailed comparison of observations to individual (hyper-/ultra-compact) \ion{H}{II} regions around forming stars in the simulation, for which the evolution is much more stochastic. 

Whilst the comparisons between our observations and the simulations of \citetalias{2010ApJ...711.1017P} provide a good foundation for further analysis, it would be beneficial to make additional comparisons with a suite of simulations. Such simulations could be fine-tuned to match the observations and model the formation and evolution of \ion{H}{II} regions with the same global properties and in the same environment, with a range of different initial conditions. However, this is outside the scope of the paper.

The hydrodynamical simulations of \citetalias{2010ApJ...711.1017P} describe the gravitational collapse of a rotating molecular gas cloud to form a cluster of massive stars. The 3D model takes into account heating by ionising and non-ionising radiation using an adapted \textsc{flash} code \citep{2000ApJS..131..273F}. 
The synthetic RRL maps of the simulation data are produced using \textsc{radmc-3d} \citep{2012ascl.soft02015D} as described in \citet{2012MNRAS.425.2352P}. Both local thermodynamic equilibrium (LTE) and non-LTE simulations of H70$\upalpha$ emission are run for a total of $0.75$ Myr. We use RRL data corresponding to 730.4, 739.2, 746.3, 715.3 and 724.7 kyr for which an ionised bubble has already emerged.

Summarising \citet{2010ApJ...711.1017P,2010ApJ...725..134P}, the simulated box has a diameter of $3.89$ pc with a resolution of $98$ AU. The initial cloud mass is $1000$ M$_{\sun}$, with an initial temperature of $30$ K, and core density $3.85 \times 10^{3}$ cm$^{-3}$. Beyond the flat inner region of the cloud ($0.5$ pc radius), the density drops as $r^{-3/2}$. The initial velocities are pure solid-body rotation without turbulence, with angular velocity $1.5 \times 10^{-14}$ s$^{-1}$, and a ratio of rotational energy to gravitational energy, $\beta = 0.05$.
Sink particles (of radius 590 AU) form when the local density exceeds the critical density, $\rho_{\text{crit}} = 2.12 \times 10^{8}$ cm$^{-3}$ and the surrounding region around the sink particle, $r_{\text{sink}} = 590$ AU, is gravitationally bound and collapsing. The sink particles accrete overdense gas that is gravitationally bound, above the threshold density, and within an accretion radius. The accretion rate varies with time and is different for each sink particle.
Within the first 10$^5$ years since the formation of the first sink particle, the original star has accreted 8 M$_{\sun}$ and many new sink particles have formed. In the next $3 \times 10^{5}$ yr, the initial three sink particles have masses of $10$-$20$ M$_{\sun}$ and no star reaches a mass greater than $25$ M$_{\sun}$ overall.

Figures \ref{bubbles} and \ref{finalbubbles} show density slices of the simulation, for the last $100$ kyr. The vectors indicate velocity and the white points represent sink particles. Figure \ref{bubbles} shows four snapshots equivalent to the initial evolutionary stages before the \ion{H}{ii} region forms, occurring at $614.0$, $624.3$, $652.7$ and $668.2$ kyr. Initially, the cloud looks square as a consequence of not including turbulence in the initial conditions and the use of a grid-based code. The central rarefaction, and surrounding dense, ring-like structure may be a result of the cloud undergoing a rotational bounce (i.e. when the core --- formed after the collapse of a rotating cloud --- continually accretes from the envelope and then expands due to rotation and the increased gas pressure gradient, resulting in a ring at the cloud's centre; \citealt{2003MNRAS.340...91C}). 
Figure \ref{finalbubbles} shows the snapshots at the final stages of the run after the \ion{H}{ii} region has formed, at $730.4$, $739.3$, and $746.3$ kyr. The thin white border encloses a region that has surpassed a 90\% ionisation fraction.

\subsection{Observations and Simulations Compared}
It is difficult to make an exact comparison between the observations and simulations, since we cannot observe the gas cloud of G316.81--0.06 at its initial stages. 
\citet{1996A&AS..118..191J} calculated the density and cloud mass of G316.81--0.06 in their multi-transition CS study, finding a mass of 1060 M$_{\sun}$ and number density $10^{4}$ cm$^{-3}$ which is in excellent agreement with the simulations\footnote{The author also identifies a velocity gradient across the CS core. Unfortunately, the value is not specified.}.

We can also estimate the mass and size of the region from the IRDC. 
\citet{2017MNRAS.470.1462L} calculated the mass of the IRDC in G316.752+0.004, a region which encompasses G316.81--0.06. They found a mass of $1.5 \times 10^4$ M$_{\sun}$, however, their distance to the IRDC is highly uncertain. Of the two distances they derive, they adopt the farther distance of $9.8$ kpc as opposed to the nearer distance of $2.6$ kpc (which is also the distance used here). Using the nearer distance estimate, the mass of the IRDC is $\sim1150$ M$_{\sun}$ which is also in agreement with the initial molecular mass of the simulated cloud of \citetalias{2010ApJ...711.1017P} ($1000$ M$_{\sun}$). 
Assuming a distance of $2.6$ kpc, the masses of the observed and simulated clouds are similar. 

The sizes of the observed and simulated regions are also similar. The area encompassing both \ion{H}{ii} regions within G316.81--0.06 is $\sim0.9$ pc in diameter, although we realise that the IRDC from which the \ion{H}{ii} regions formed is certainly larger than this. The initial central condensed structure at 500 kyr of the simulations is $1.3$ pc in diameter.
From this, we infer that the density of the observations and simulations will also be on the same order of magnitude, in agreement with the aforementioned result of \citet{1996A&AS..118..191J}. Given the similarity between the mass and size of the \ion{H}{ii} regions we conclude that it is reasonable to compare the observations to the simulations (bearing in mind the caveats in \S\, \ref{sims}). 


\section{Data Analysis}
\label{method}
The data analysis was performed using the Common Astronomy Software Applications (\textsc{casa}; \citealt{2007ASPC..376..127M}) package and the Semi-automated multi-COmponent Universal Spectral-line fitting Engine (\textsc{scouse}; \citealt{2016MNRAS.457.2675H}). \textsc{casa} was used to calculate 2$^\text{nd}$ moment maps (velocity dispersion, $\upsigma$), and Gaussians were fit to the spectra using \textsc{scouse} in order to determine centroid velocity (v$_0$). 
The input parameters for \textsc{scouse} fitting are found in Table \ref{tab:scouse}, according to \citet{2016MNRAS.457.2675H}. 

\begin{table}
\caption{\textsc{scouse} input. Parameter names \newline according to \citet{2016MNRAS.457.2675H}.}
\label{tab:scouse}
\begin{tabular}{lcc}
\hline
Parameter & Observations & Simulations \\
\hline
$R_{\text{SAA}}$ & 0.$\degr$001 & 0.$\degr$0003 \\
RMS (K) & 0.02 & 0.06 \\
$\sigma_{\text{rms}}$ (K) & 3.0 & 3.0 \\
$T_1$ & 5.0 & 5.0 \\
$T_2$ & 2.5 & 2.5 \\
$T_3$ & 2.5 & 1.7 \\
$T_4$ & 1.0 & 1.00\\
$T_5$ & 0.5 & 0.5\\
$v_{\text{res}}$ (km s$^{-1}$) & 0.5 & 1.56\\
\hline
\end{tabular}
\end{table}

\subsection{Observations} 
We used the H70$\upalpha$ RRL spectra taken and reduced by \citet{2009MNRAS.399..861L} for the data analysis. With \textsc{casa}, the 2$^\text{nd}$ moment map was created between velocities -9.3 and -70.7 km s$^{-1}$ including only pixels above 26 mJy beam$^{-1}$ in order to optimally exclude the weaker, second velocity component (see below). 

We found that towards the south-west of Region 1, the spectra contain an additional component which is broader (by 14.8\%) 
and less intense (by 73.5\%) 
than the primary component. Inspection of the data cubes shows that this emission is offset both in velocity and spatially, and we conclude is unassociated with the ionised gas of Region 1. 

\begin{figure}
\centering
\includegraphics[width=0.49\textwidth]{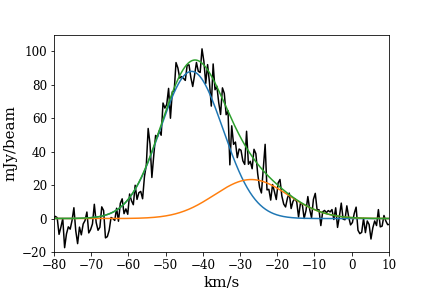}
\caption{H70$\upalpha$ spectra of the observational data (Region 1). With \textsc{scouse}, a two-component fit was applied to prevent contamination in further analysis. The primary component (blue) and secondary component (orange) are each fitted by a Gaussian. The combined fit is shown in green.}
\label{two_comp}
\end{figure}

Figure \ref{two_comp} shows Gaussian fits to both components, identified with \textsc{scouse}. Where possible, the contribution of the secondary component was excluded from further analysis, as we are interested in Region 1. At locations where the secondary component is much weaker, it became difficult to distinguish between the two components. This means that we cannot create FWHM maps reliably, and that the results from the area covering the lowest third of Region 1 must be treated with caution. 

\subsection{Simulations}
In order to compare the simulations and observations more robustly, the units of the H70$\upalpha$ synthetic data have been transformed to be consistent with the observations. Intensity was converted from erg s$^{-1}$ cm$^{-2}$ Hz$^{-1}$ ster$^{-1}$ to Jy beam$^{-1}$; physical size converted to an angular size using the distance to G316.81--0.06 (2.6 kpc); and frequency converted to velocity. Using \textsc{casa}, the continuum was subtracted using \texttt{imcontsub} with the line-free channels: $\leq 6$ and $\geq 54$ (LTE); $\leq 10$ and $\geq 54$ (non-LTE). 

A major difference between the LTE and non-LTE simulations were narrow absorption lines (LTE) and very bright, compact, and narrow emission lines (non-LTE) which perhaps emulate real maser emission. In non-LTE conditions, RRLs may undergo maser amplification when the line optical depth is negative and its absolute value is greater than the optical depth of the free-free emission (\citealt{2002ASSL..282.....G} and references therein).
The narrow emission dominates over the broad RRL emission, and appears almost as a delta function which prevents \textsc{scouse} from being able to fit the non-LTE simulations. Therefore, a mask was applied to remove the majority of the narrow emission; every value greater than 10 mJy beam$^{-1}$ was replaced with the average of the points either side. 

\begin{figure}
\centering
\includegraphics[width=0.49\textwidth]{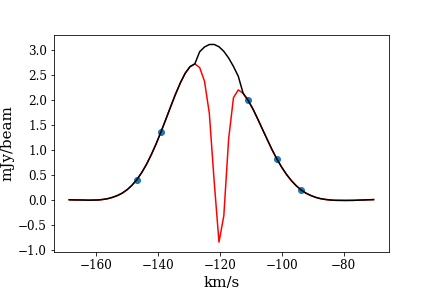}
\caption{RANSAC example applied to the LTE synthetic data. The narrow absorption lines prevented successful fitting with \textsc{scouse} so they were replaced. The original spectrum (red), the new spectrum (black), and the blue circles represent the five points used to make a Gaussian fit. In this example, this fit was chosen to be the best out of three hundred iterations.}
\label{ransac}
\end{figure}

The narrow absorption lines (LTE) also made \textsc{scouse} fitting difficult. Significant portions of the broad RRL emission were often missing, making it challenging to fit the overall structure. Therefore, the absorption lines were removed via a RANdom SAmple Consensus (RANSAC; \citealt{Fischler:1981:RSC:358669.358692}) method\footnote{For consistency, RANSAC was also applied to the non-LTE data after the narrow emission lines were removed.}. RANSAC iteratively estimates the parameters of a mathematical model from a set of data and excludes the effects of outliers.
In our case, we selected five points at random (blue circles) along each spectrum (red) to make a Gaussian fit (Figure \ref{ransac}). Using RANSAC, the best fit is the fit with the most inliers (points with a residual error of less than 5\%) out of three hundred iterations. Values of the original spectrum which lay outside the threshold (5\%) are replaced by the values from the new fit so as not to entirely eradicate the original data. With this method we were able to successfully remove the narrow absorption lines without distorting the data, so that we could then proceed with the \textsc{scouse} fitting. This was only successful for the last three timesteps (730.4, 739.2, and 746.3 kyr). At earlier times (715.3 and 724.7 kyr), for which synthetic H70$\upalpha$ data are also available, the LTE absorption lines are too wide to be accurately removed via the RANSAC method, and thus cannot be successfully fit by \textsc{scouse}. 

Finally, Gaussian smoothing was applied to both the LTE and non-LTE synthetic data with a beam size of 2.5 arcsec, using the \texttt{imsmooth} tool in \textsc{casa}. This is the largest beam size possible while still being able to resolve the overall kinematic structure. 

\section{Results}
\label{results}

\begin{figure}
	\includegraphics[width=0.49\textwidth]{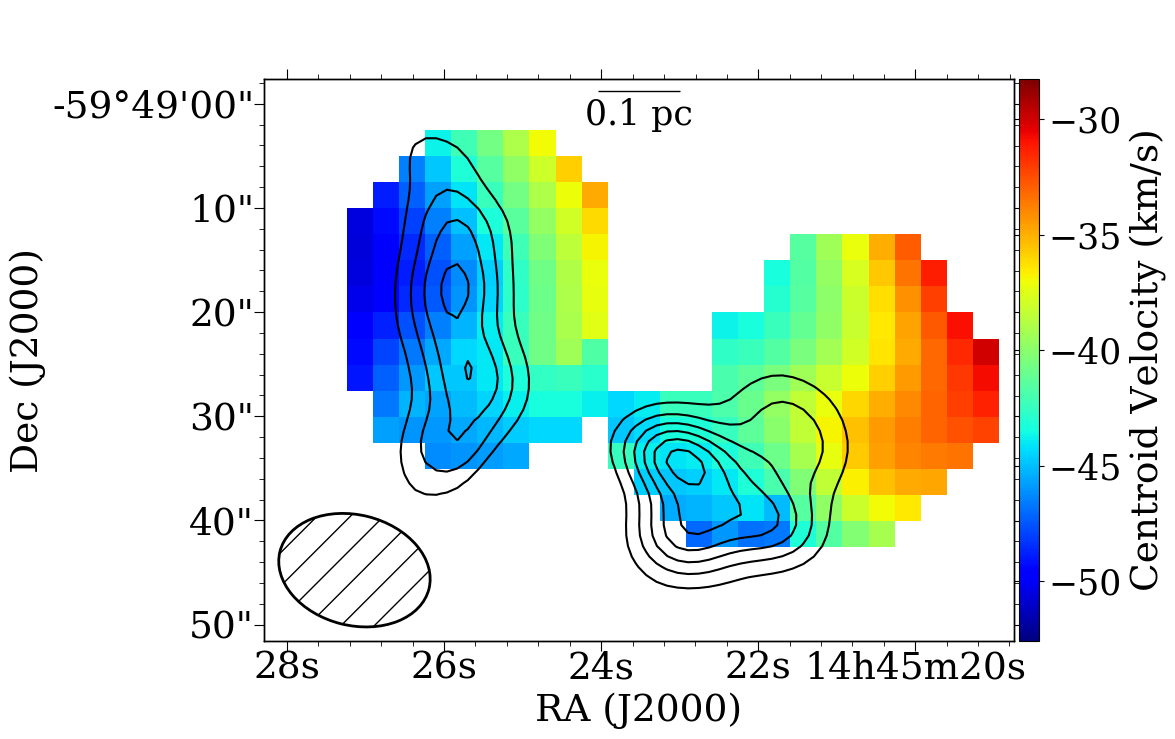}
	\includegraphics[width=0.49\textwidth]{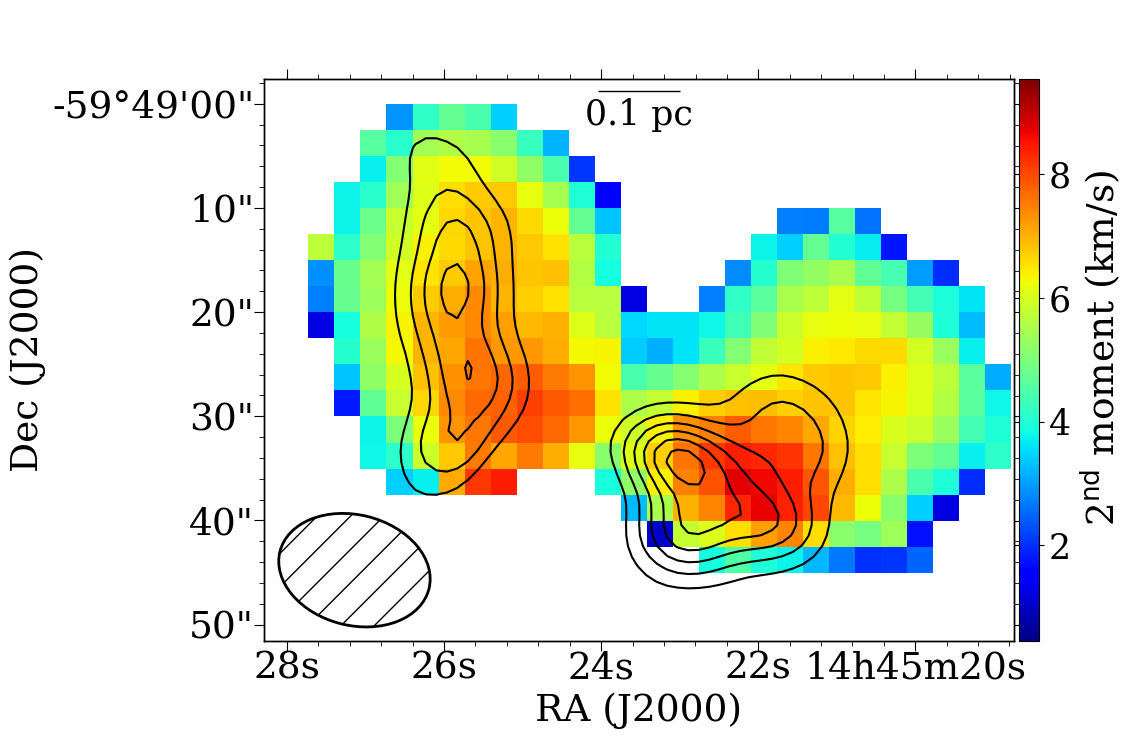}
	\caption{Maps of the two \ion{H}{ii} regions within G316.81--0.06: Region 1 (left) and Region 2 (right). Top: centroid velocity map (\textsc{scouse}); bottom: 2$^\text{nd}$ moment map (\textsc{casa}). The beam is shown at the bottom left of each map. Contours of the 35-GHz continuum are overlaid in black \citep{2009MNRAS.399..861L}.}
    \label{fig:obs}
\end{figure}

\begin{figure*}
	\centering
	\includegraphics[width=0.43\textwidth]{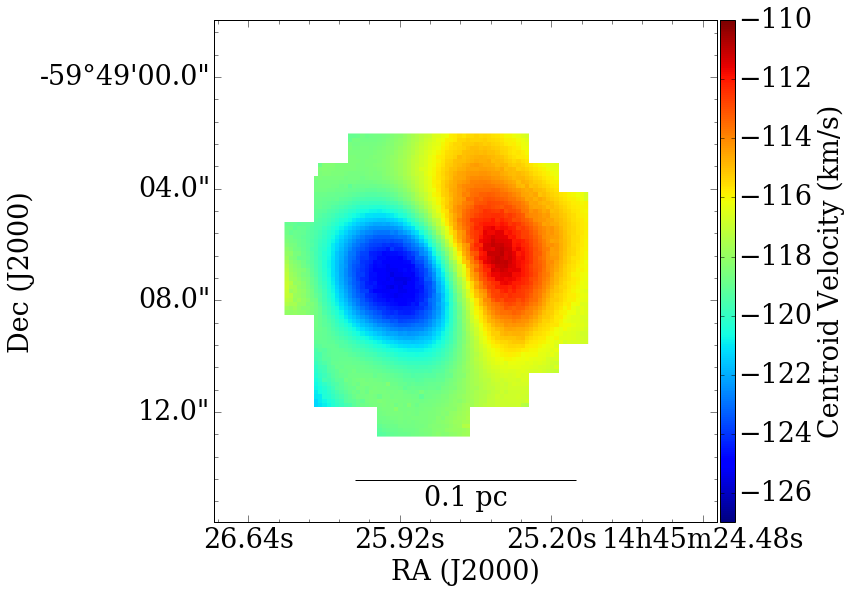}
    \includegraphics[width=0.43\textwidth]{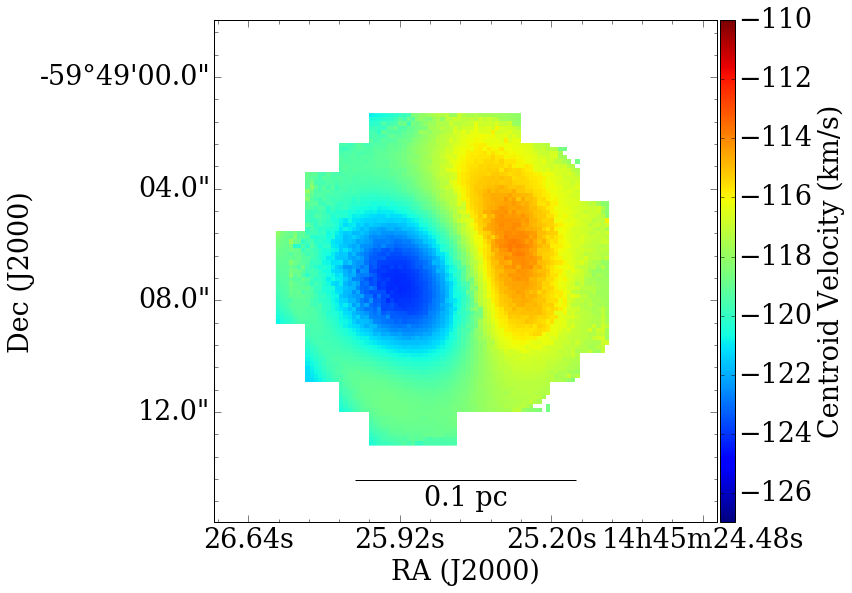}
    \includegraphics[width=0.43\textwidth]{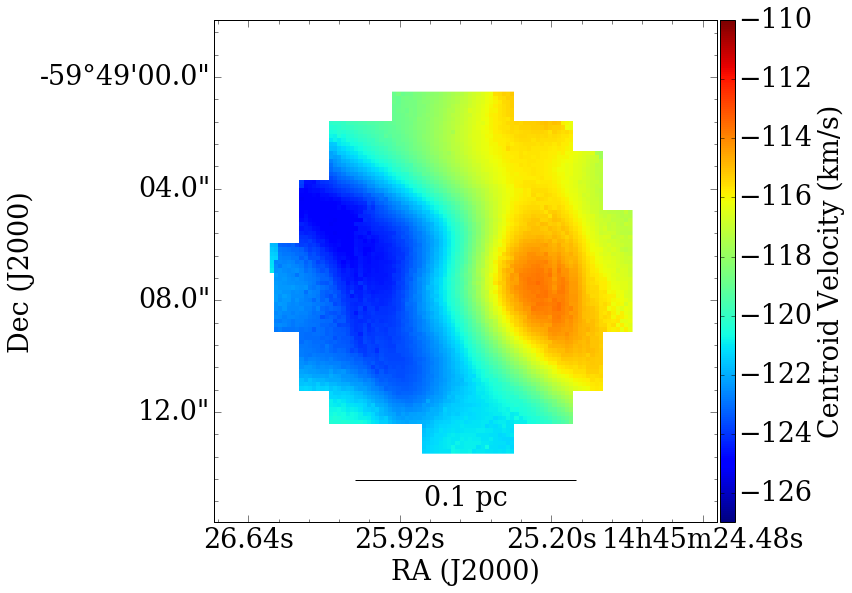}
	\includegraphics[width=0.43\textwidth]{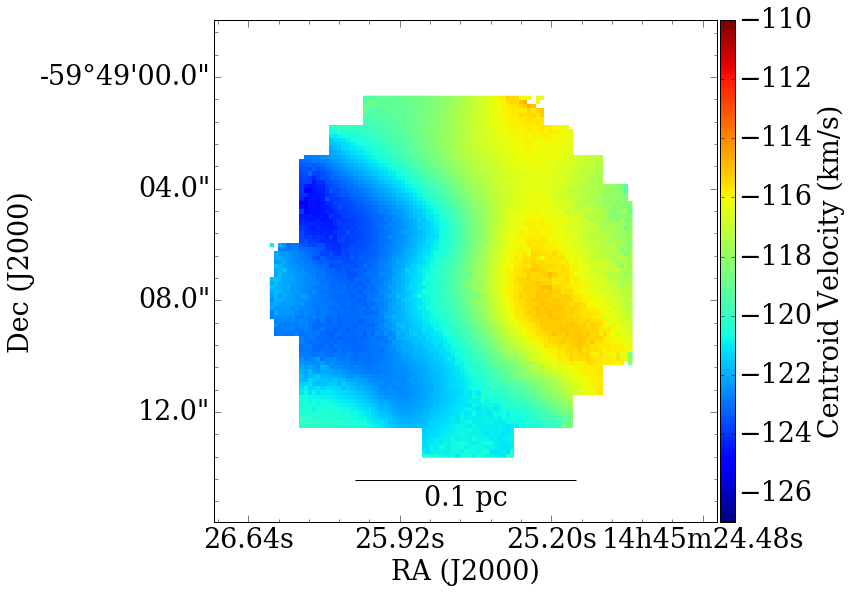}
    \includegraphics[width=0.43\textwidth]{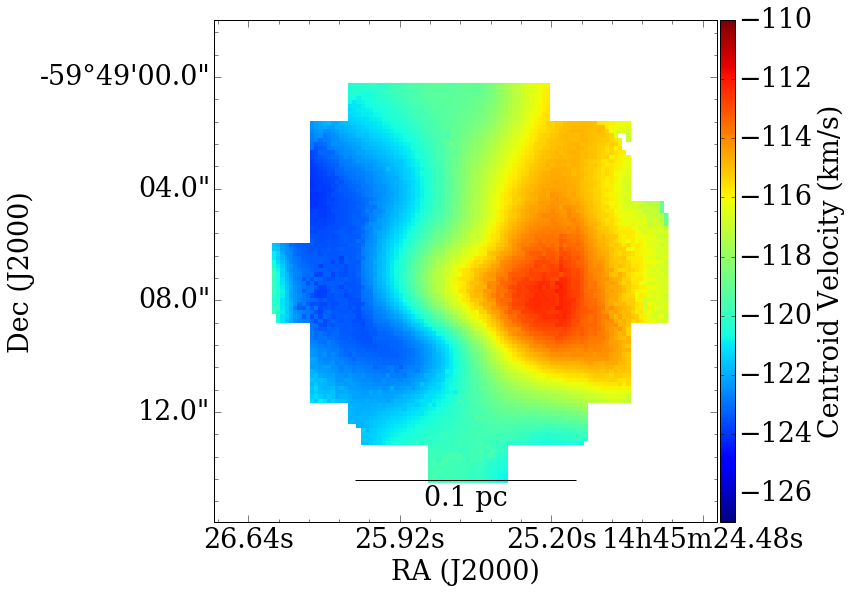}
    \includegraphics[width=0.43\textwidth]{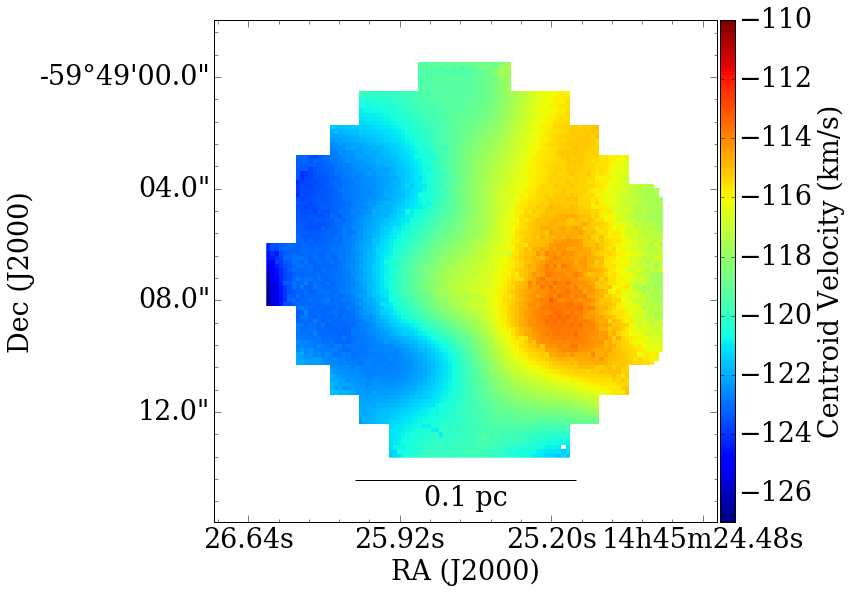}
	\caption{\textsc{scouse} outputted centroid velocity of the simulated H70$\upalpha$ data. Left: LTE; right: non-LTE, at ages of 730.4, 739.3, and 746.3 kyr increasing from top to bottom.}
    \label{fig:sims_v}
\end{figure*}

\begin{figure*}
	\centering
	\includegraphics[width=0.43\textwidth]{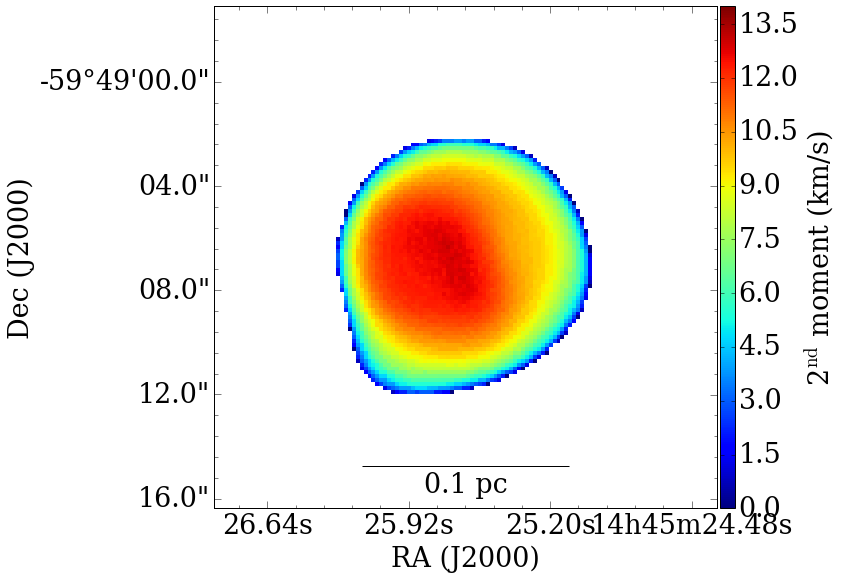}
    \includegraphics[width=0.43\textwidth]{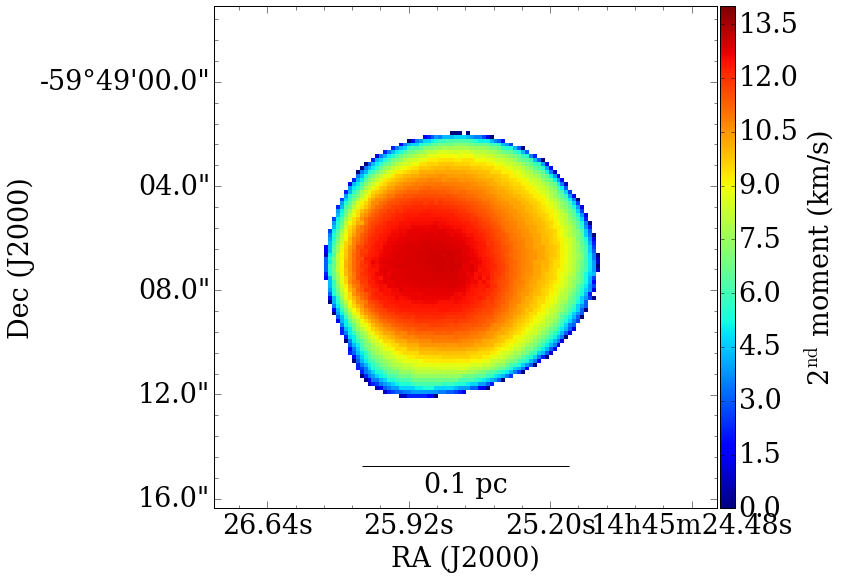}
	\includegraphics[width=0.43\textwidth]{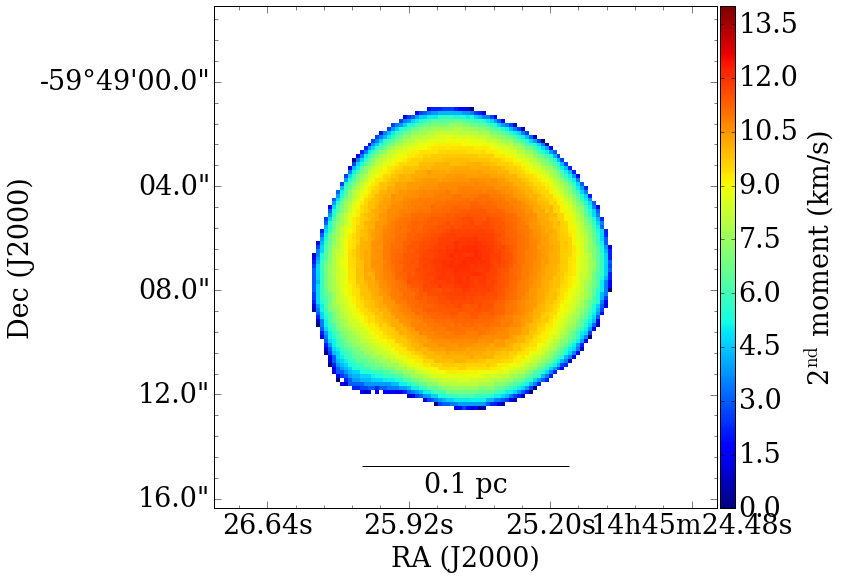}
    \includegraphics[width=0.43\textwidth]{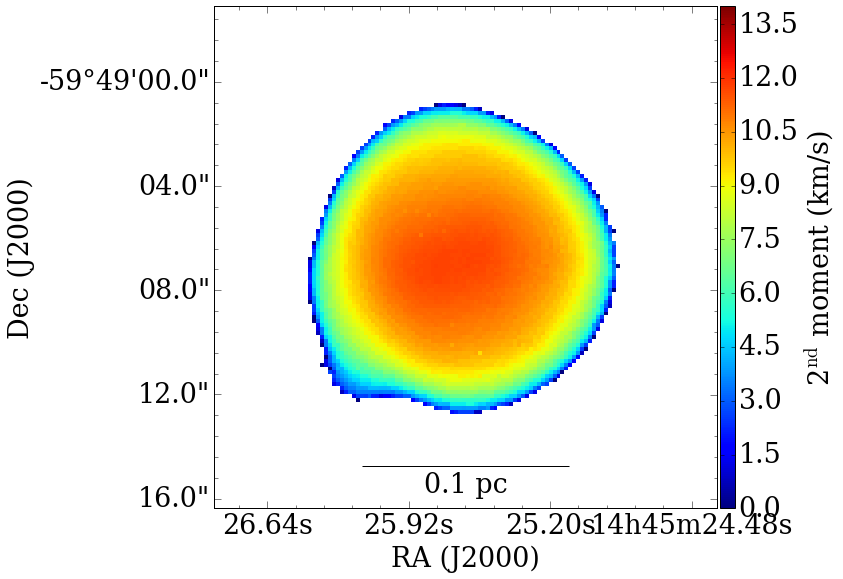}
    \includegraphics[width=0.43\textwidth]{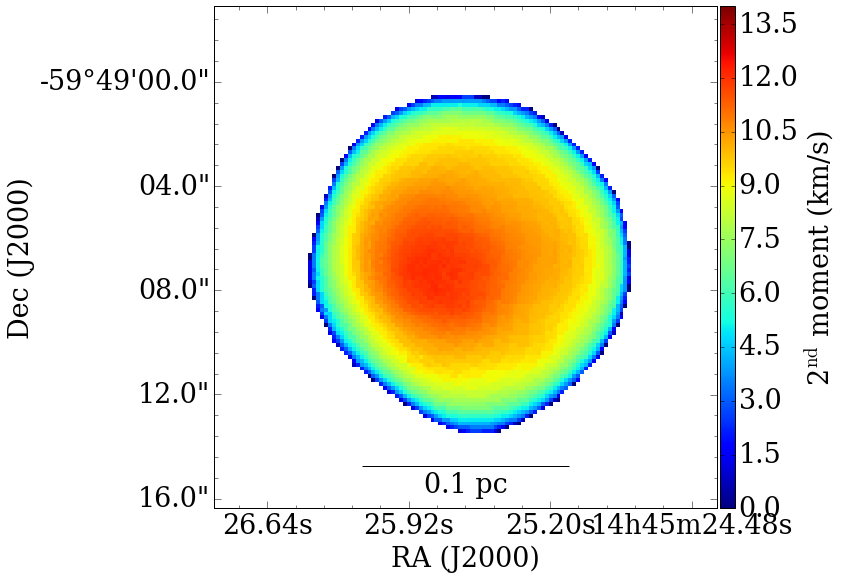}
    \includegraphics[width=0.43\textwidth]{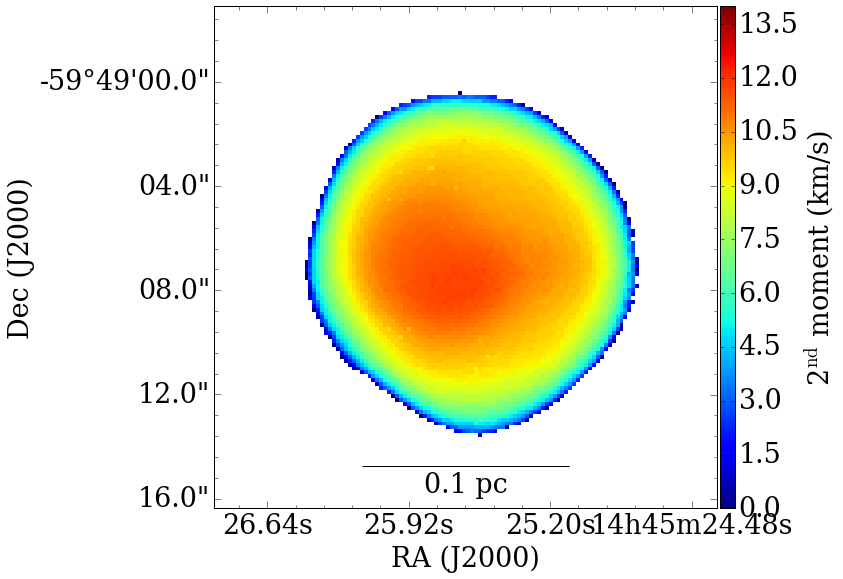}
	\caption{2$^\text{nd}$ moment maps of the simulated H70$\upalpha$ data outputted by \textsc{casa}, showing the velocity dispersion. Left: LTE; right: non-LTE, at ages of 730.4, 739.3, and 746.3 kyr increasing from top to bottom.}
    \label{sims_mom2}
\end{figure*}

Table \ref{tab:results} contains the ranges in velocity (v$_0$(max)-v$_0$(min)), velocity gradients and maximum velocity dispersions for both the observational and synthetic H70$\upalpha$ RRL data. 
Where the observations are concerned, Region 1 is the focus of the study, for it is the youngest \ion{H}{II} region (Figure \ref{fig:obs}). We note that the velocity gradient measured is what we observe along our line of sight, and does not take into account any inclination that may be present. For the simulations (both LTE and non-LTE), we show the results of the final three ages: 730.4, 739.2, and 746.3 kyr in Figures \ref{fig:sims_v} and \ref{sims_mom2}.

\subsection{Observed ionised gas kinematics}

Figure \ref{fig:obs} contains the H70$\upalpha$ centroid velocity and 2$^\text{nd}$ moment maps for the two \ion{H}{II} regions in G316.81--0.06. 
Several velocity gradients across each region are visible in the centroid velocity map, and we focus on the younger \ion{H}{II} region; Region 1 (left). Region 1 shows a velocity gradient roughly east-west (across $\sim0.33$ pc) in addition to a less steep gradient north-south (across $\sim0.42$ pc) aligned with the elongation of the 35-GHz continuum and `green fuzzy'. 
The 2$^\text{nd}$ moment map shows that in both \ion{H}{ii} regions $\upsigma$ increases towards the centre, and is highest towards the southern end of each region.

\subsection{Simulated ionised gas kinematics} 

Figures \ref{fig:sims_v} and \ref{sims_mom2} show the centroid velocity and 2$^\text{nd}$ moment maps of the LTE and non-LTE synthetic H70$\upalpha$ data, at the final three ages 730.4, 739.2, and 746.3 kyr. The density slices (Figures \ref{bubbles} and \ref{finalbubbles}) look down on the stars in the xy-plane, i.e. along the outflow axis. For the synthetic H70$\upalpha$ data, we chose a projection perpendicular to the outflow plane (oriented along the xz-plane), to compare with the observations. Although the inclination of the observed Region 1 outflow is not known; Figure \ref{fig:image} shows it is clearly closer to perpendicular than along the line-of-sight. Any inclination will introduce a change in observed velocity motions of order sin($\theta$), where $\theta$ is the angle of inclination. 

The velocity structure of the simulated ionised gas on 0.05 pc scales changes insignificantly between the different time steps for either the non-LTE or LTE synthetic maps. Given the similar kinematic structure between the non-LTE and LTE synthetic maps, we conclude non-LTE effects are not important for our analysis and focus on the LTE maps from here on. As this kinematic structure is a robust feature of the simulations, it seems reasonable to compare to the observed ionised gas kinematics.  

The morphology of the centroid velocity and 2$^\text{nd}$ moment maps are similar to the observations; velocity gradients are oriented roughly east-west and velocity dispersion increases towards the centre. A significant difference is that the simulated \ion{H}{ii} region is smaller ($\sim0.15$ pc versus $\sim0.33$ pc), potentially resulting in the steeper velocity gradients compared to the observations, due to the conservation of angular momentum. 


\begin{table*}
\caption{Values for the range in centroid velocity, v$_0$, velocity gradient, $\nabla$v$_0$, and maximum velocity dispersion, $\upsigma_{\text{max}}$, for both \ion{H}{ii} regions in G316.81--0.06, in addition to the LTE and non-LTE simulations of \citetalias{2010ApJ...711.1017P} for the final three time-steps.} 
\label{tab:results}
\begin{tabular}{lcccccccc}
\hline
 &Region 1  &Region 1 & LTE&LTE&LTE & non-LTE&non-LTE&non-LTE \\
Time (kyr) &(E-W)&(N-S)& 730.4 & 739.2 & 746.3 & 730.4 & 739.2 & 746.3\\
\hline
v$_0$ range (km s$^{-1}$) & 15.78$\pm$0.45 & 5.14$\pm$1.11 & 14.59$\pm$0.01& 11.64$\pm$0.01& 12.04$\pm$0.03& 10.49$\pm$0.05& 10.40$\pm$0.12 & 13.54$\pm$0.12 \\
$\nabla$v$_0$ (\kmspc) & 47.81$\pm$3.21 & 12.23$\pm$2.70 & 97.29$\pm$12.97 & 77.64$\pm$10.35 & 80.25$\pm$10.70 & 69.91$\pm$9.33 & 69.35$\pm$9.28 & 90.28$\pm$12.07 \\
$\upsigma_{\text{max}}$ (km s$^{-1}$) &8.1 & 8.1 & 13.1 & 12.1 & 12.1 & 13.0 & 11.8 & 11.8\\
\hline
\end{tabular}
\end{table*}

\section{Discussion}
\label{dis}  

As introduced in \S\, \ref{intro}, prior literature looking at the ionised gas kinematics of \ion{H}{II} regions has primarily focused on expansion, accretion, and outflows. There are, however, a small number of \ion{H}{II} regions in the literature which display velocity gradients perpendicular to the outflow axis. For these regions, a common interpretation is that rotation in some form is contributing to the velocity structure.
In \S\, \ref{examples} we summarise previous observations put forward as evidence that rotation is playing a role in shaping the velocity gradient. In \S\, \ref{sim_rotation} we turn to the \citetalias{2010ApJ...711.1017P} simulations to try and uncover the origin of the velocity gradient perpendicular to the outflow axis. Section \ref{rotation} refers back to Region 1, discussing whether the velocity structure signifies rotation and what can be inferred with relation to feedback.

\subsection
[{Postulated evidence for rotation in observed H II regions}]
{Postulated evidence for rotation in observed H\,{\sevensize II} regions}
\label{examples}

\begin{description}
\item \textit{\textbf{G34.3+0.2C}}.
Although a cometary \ion{H}{II} region, a remarkably strong velocity gradient of $\sim 338$ \kmspc \, has been detected in the H$76\upalpha$ RRL, perpendicular to the axis of symmetry. \citet{1986ApJ...309..553G} infer that this could be caused by a circumstellar disc which formed from the collapse of a rotating protostellar cloud. They suggest that the angular velocity of the cloud was a result of Galactic rotation, and find that the angle of rotation roughly aligns with that of the Galactic plane. \citet{1994ApJ...432..648G} refute this since the surrounding molecular material appears to rotate opposite to the ionised gas. Instead, they suggest that stellar winds from two nearby sources have interacted with the ionised gas to give the observed velocity profile.

\item \textit{\textbf{W49A}}. \citet{1977ApJ...212..664M} present a low spatial resolution study of W49A, and find a velocity range of a few km s$^{-1}$ across the bipolar \ion{H}{II} region. In comparison to the velocities of two massive molecular clouds either side of the \ion{H}{II} region, they conclude that the ionised gas rotates in the middle of them, as the molecular clouds revolve about one another. With higher spatial resolution, \citet{1987Sci...238.1550W} find a 2-pc ring containing at least ten separate \ion{H}{II} regions which they claim is rotating about $5\times10^4$ M$_{\sun}$ of material. They derive an angular velocity of 14.4 \kmspc. With even higher spatial resolution, \citet{1997ApJ...482..307D} find 45 distinct continuum sources, and that the \ion{UCH}{II} regions within the ring do not appear to have ordered motions. However, one \ion{UCH}{II} region in particular (W49A/DD), shows a north-south velocity gradient of a few km s$^{-1}$ which \citet{1997ApJ...482..307D} claim may be caused by the rotation of the ionised gas.

\item \textbf{\textit{K3-50A}}. The bipolar \ion{H}{II} region, K3-50A, shows a steep velocity gradient ($\sim150$ \kmspc) along the axis of continuum emission, indicating the presence of ionised outflows \citep{1994ApJ...428..670D}. There also appears to be an unmentioned perpendicular velocity gradient across the region which we estimate to be $\sim30$ \kmspc\, (see Figure 5a of \citealt{1994ApJ...428..670D}). Further detailed comparisons with the molecular disc have been made \citep{1997ApJ...477..738H} in addition to polarimetry studies \citep{2015MNRAS.453.2622B}. This has provided a unique insight to the influence of magnetic fields and allowed for the construction of a detailed 3D model.

\item \textbf{\textit{NGC 6334A}}.
The velocity gradient of this bipolar \ion{H}{II} region was first detected by \citep{1988BAAS...20.1031R}. \citet{1995ApJ...447..220D} reconfirmed this, finding a gradient of $\sim$75 \kmspc. They inferred that the signature can be attributed to rotation of the ionised gas, originating from a circumstellar disc. They derived a core Keplerian mass of $\sim$200 M$_{\sun}$. 

It has also been noted that NGC 6334A, K3-50A, and W49A/A are all alike in terms of their bipolar morphology and the possible presence of  ionised outflows \citep{1997ApJ...482..307D}. 
\end{description}

\subsection{Origin of the ionised gas velocity structure in the P10 simulations}
\label{sim_rotation}

Since it is difficult for models to take into account all of the different physical mechanisms involved with the ionisation process, one common simplification is to use static high-mass stars. Such simple analytic models tend to show that ionisation occurs isotropically, for a homogeneous surrounding medium, resulting in no velocity gradient. However, in most simulations it is clear that stars are in motion with respect to each other and the surrounding gas. 

In the \citetalias{2010ApJ...711.1017P} simulations this motion results in a preferred direction of ionisation, downstream of the stellar orbit. We present a simple cartoon (Figure \ref{cartoons}) based on a qualitative examination of the simulated density vector maps (Figures \ref{bubbles} and \ref{finalbubbles}). We find that this can explain the red- and blue-shifted spectra of the observed RRL profile. The cartoon illustrates the evolutionary sequence beginning at the formation of the initial molecular cloud up to the formation of the \ion{H}{ii} region, explained in more detail by the following: 

\begin{figure*}
\centering
\includegraphics[width=0.9\textwidth]{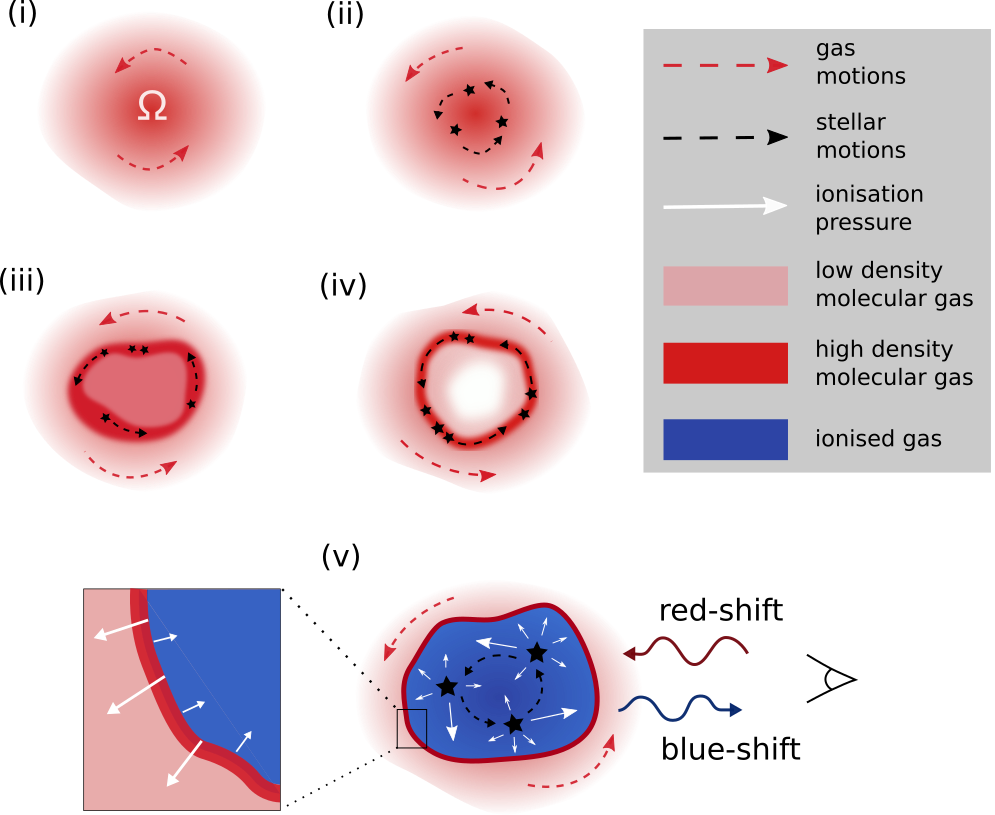}
\caption{A cartoon illustrating the kinematic evolution of a young \ion{H}{ii} region. (i) The molecular gas cloud forms with some initial net angular momentum ($\upOmega$, red dashed arrows), with increasing density towards the centre of the cloud (pink). (ii) Stars (black) form at the centre of the molecular cloud when the critical density is surpassed, then drift outwards. The stars orbit about the cloud's centre, tracing the angular momentum of the cloud (black dashed arrows). (iii) A ring (solid red) forms as a result of rotational bounce \citep{2003MNRAS.340...91C}. Stars continually gain mass and accrete material which gravitationally collects about the stars. The higher density ring initiates the formation of new stars. (iv) The centre becomes rarefied, meanwhile newly ionised material rapidly recombines (also known as flickering). (v) Ionisation dominates over recombination resulting in an \ion{H}{ii} region (blue). Ionisation is strongest to the front of the star's path where the pressure is lowest (white arrows). This appears as blue-shifted spectra when the star travels and ionises towards us, and red-shifted when the star travels and ionises away from the observer.
Molecular material collects about the edge of the ionised region as the bubble expands (thick red solid line).}
\label{cartoons}
\end{figure*}

\begin{enumerate}
\item The initial molecular gas cloud has some net angular momentum ($\upOmega$, red dashed arrows), with increasing density towards the centre. 
\item Once the local critical density of the gas is surpassed, stars (black) form with a high star formation efficiency at the centre of the cloud. The first star forms at the centre of the potential well then quickly drifts outwards, soon followed by the formation of more stars (on timescales of kyr). These stars all immediately begin to trace the rotation of its natal cloud, about the centre of mass (black dashed arrows).
\item The central region starts to become rarefied and a ring-like structure appears (solid red)\footnote{We note in passing the similarity to the ring of \ion{H}{II} regions in W49A (\S\, \ref{examples}).}, likely a result of rotational bounce \citep{2003MNRAS.340...91C}. Simultaneously, material accumulates about the stars, and new stars form within the dense material, and in general continue to trace the rotation of the molecular cloud. The first star makes approximately one complete revolution until the simulation ends (across $\sim120$ kyr), taking into account that the stars are continually moving outwards as they orbit.
\item  In the simulations, the rarefied centre is made up of inhomogeneous regions of lower density and lower pressure (shown as solid white for simplicity). Newly ionised material rapidly recombines (also known as flickering; e.g. \citetalias{2010ApJ...711.1017P}; \citealt{2011MNRAS.416.1033G}). The stars continue to orbit about the centre of mass, and also interact with each other, some getting flung outside of the cloud. For a detailed description of stellar cluster formation in this simulation see \citet{2010ApJ...725..134P}. 
\item Multiple high-mass ionising stars create one large ionised bubble (solid blue) also containing lower-mass stars. The thermal pressure created by ionisation heating drives the expansion of the \ion{H}{II} region (white arrows), sweeping the surrounding neutral material into a dense shell (thick solid red line). The thermal pressure of the ionised gas is two orders of magnitude higher than in the molecular gas, and thus the pressure gradient term of the Euler equation dominates over the advection term at the \ion{H}{II} region boundary. Hence, the ionised gas does not trace the rotation of the molecular gas directly. The stars act as mediators, inheriting their angular momentum from the molecular gas out of which they formed, and then create angular momentum in the ionised gas via a different mechanism (described in more detail below). This is shown by the magnitudes and directions of the velocity arrows of the ionised gas (Figure \ref{finalbubbles}). If the ionised gas
was put in rotation by the surrounding molecular gas, then the arrows on either side of the \ion{H}{II} region boundary
should always point in the same direction, which is clearly not the case. Furthermore, the varying length of the arrows within the
\ion{H}{II} region provides evidence for strong dynamical processes inside the \ion{H}{II} region that would destroy any such
coherent velocity pattern coming from the boundary.

In fact, it is these dynamical processes that generate the rotational signature in the ionised gas as follows.
Typically, models use an idealised scenario whereby the star is static and ionises isotropically (e.g. \citealt{1978ppim.book.....S}). 
However, in the frame of reference where the star is stationary, consider that upstream of the star's path, the gas flow in the cloud is in opposition to the direction of ionisation. This inhibits expansion of the ionised gas, as it is continually replenished by neutral material and recombines. Whereas downstream of the star's path, the neutral material travels with the direction of ionisation; the pressure is lowest ahead of the star in comparison to all other directions. Therefore, the expansion occurs predominantly in front of the star as it orbits, i.e. the path of least resistance. 
The velocity of the ionised gas traces the orbit of the stars and gas and hence, we observe red- and blue-shifted spectra along our line-of-sight in the ionised gas of the simulation.
\end{enumerate}

\subsection
[{G316.81--0.06: a rotating H II region?}]
{G316.81--0.06: a rotating H\,{\sevensize II} region?}
\label{rotation}

We now investigate the possible origin of the velocity structure in G316.81--0.06.
If it is solely due to outflow and/or expansion of ionised gas, the bipolar \ion{H}{II} region would need to be significantly inclined. Since we clearly observe the elongated 35-GHz continuum and `green fuzzy' along the axis of bipolarity, in addition to the shallow velocity gradient north-south, we infer that Region 1 is close to edge-on, i.e. 
the line of sight is primarily along the disc plane. While we cannot rule out that the observed velocity structure is caused only by a combination of expansion and outflow, we could not construct a simple model that explains the velocity structure with only these mechanisms. 
Therefore, we believe that rotation is the most likely explanation for the velocity gradients.

This is further supported when we compare the ionised gas kinematics in the simulations of \citetalias{2010ApJ...711.1017P} to the observations of Region 1. As shown in \S\, \ref{results}, the morphology is similar, both in terms of centroid velocity and the 2$^\text{nd}$ moment maps. We also find that the velocity gradient in the simulations is around a factor of two higher than that of Region 1. This may be due to the simulated \ion{H}{II} regions being approximately a factor of two smaller ($\sim0.15$ pc as opposed to $\sim0.33$ pc) and that the inclination of Region 1 may be non-zero.

Further work is needed to explore the significance of rotating gas as opposed to non-rotating gas. In the absence of simulations fine-tuned to match the observations, or higher resolution observations to measure the velocity/proper motions of embedded stars we have reached the limit of the extent to which we can test this scenario.
However, bearing in mind the caveats discussed in \S\, \ref{sims}, we conclude that the simulations remain a useful tool to aid our understanding of the motions of the ionised gas, especially given the simulations were not fine-tuned to the observations. Moreover, the unusual velocity gradients naturally emerge from the \citetalias{2010ApJ...711.1017P} simulations which were not designed to study this effect. 

Referring back to the previous postulated explanations from \S\, \ref{examples}, the interpretation of the velocity gradient based on comparison to the \citetalias{2010ApJ...711.1017P} simulations is most similar to the scenario put forward by \citet{1986ApJ...309..553G}. An interesting prediction of this scenario is that if the initial angular momentum of the cloud is determined by Galactic rotation, the magnitude of rotation will depend on the location in the Galaxy and the orientation of the angular momentum axis with reference to the Galactic plane. 

Although not included in the \citetalias{2010ApJ...711.1017P} simulations, another possible explanation for the rotation of G316.81--0.06 may be due to the IRDC, clearly seen in Figure \ref{fig:image}. Accretion from the filament would likely induce some net angular momentum onto a central core.
Watkins et al. (in prep.) are currently studying the molecular gas kinematics of G316.81--0.06. Their results will allow for a detailed comparison between the motions of ionised and molecular gas in order to test this scenario. 

If either scenario can describe the origin of ionised gas motions in many \ion{H}{ii} regions, similar velocity gradients should also be evident in RRLs for other young (bipolar) \ion{H}{ii} regions. In order to test the scenario of rotation induced by filamentary accretion, comparative studies between the kinematics of \ion{H}{ii} regions and IRDCs are required. Galactic plane surveys, with upcoming and highly sensitive interferometers (e.g. EVLA- and SKA-pathfinders), will provide a high resolution census of all ionised regions in the Milky Way. These \ion{H}{ii} regions will be at different locations in the Galaxy, with different orientations and magnitudes of angular momentum with respect to Earth. We will have an invaluable test-bed at the earliest and poorly understood phases of star formation, allowing for the study of RRLs in \ion{H}{II} regions across a large range of ages, sizes, and morphologies.
Future high resolution observational surveys in combination with suites of numerical simulations will also further our understanding of the differing contributing feedback mechanisms at early evolutionary stages and may help to constrain different star/cluster formation scenarios.

For example, the simulations of \citetalias{2010ApJ...711.1017P} can give an idea of which feedback mechanism(s) have an important effect in G316.81--0.06. The simulations include both heating by ionising and non-ionising radiation, where the latter's only effect is to increase the Jeans mass (see discussion in \citealt{2010ApJ...725..134P}). Therefore, all dynamical feedback effects in the simulation are due to photoionisation. This may imply that ionisation pressure is the dominating feedback mechanism required for the formation of a rotating ionised gas bubble and that radiation pressure and protostellar outflows are not needed to explain the dynamical feedback. This is potentially present in all \ion{H}{ii} regions and needs to be studied further in other simulations which incorporate different feedback mechanisms.
Although the formation and evolution of galaxies will not be significantly different whether or not the outflowing gas is rotating, the potential to use the ionised gas kinematics as a tracer to identify very young \ion{H}{ii} regions represents an opportunity to understand feedback at the relatively unexplored time/size scales when the stars are just beginning to affect their surroundings on cloud scales.

\section{Summary}
\label{summary} 
We have studied a rare example of a young, bipolar \ion{H}{II} region  which shows a velocity gradient in the ionised gas, perpendicular to the bipolar continuum axis.
Through comparisons of our H70$\upalpha$ RRL observations with the synthetic data of \citetalias{2010ApJ...711.1017P}, we find that they both share a similar morphology and velocity range along the equivalent axes. 

We infer that the velocity gradient of G316.81--0.06 is the rotation of ionised gas, and that the simulations demonstrate that this rotation is a direct result of the initial net angular momentum of the natal molecular cloud. Further tests are required to deduce the origin of this angular momentum, whether it is induced by Galactic rotation, filamentary accretion, or other. If rotation is a direct result of some initial net angular momentum, this observational signature should be common and routinely observed towards other young \ion{H}{II} regions in upcoming radio surveys (e.g. SKA, SKA-pathfinders, EVLA). Further work is required to know if velocity gradients are a unique diagnostic. 

If rotation is seen to exist in other \ion{H}{II} regions, and we can uncover its true origins, this may help to parameterise the dominating feedback mechanisms at early evolutionary phases, greatly demanded by numerical studies. This should be achievable through systematic studies of many \ion{H}{II} regions, combined with comparison to a wider range of numerical simulations, likely offering a new window to this investigation.


\section*{Acknowledgements}
We would like to thank the anonymous referee for their very helpful and constructive comments. This research made use of Astropy \citep{2013A&A...558A..33A}, a community-developed core Python package for Astronomy, and APLpy \citep{2012ascl.soft08017R}, an open-source plotting package for Python. \citet{Anaconda} and \citet{Jupyter} were also used. The GLIMPSE/MIPSGAL image in Figure \ref{fig:image} was created with the help of the ESA/ESO/NASA FITS Liberator \citep{2012ascl.soft06002L}. The cartoon of Figure \ref{cartoons} was created using Inkscape \citep{Inkscape}, the free and open source vector graphics editor. Thanks are also due to Dr. Stuart Lumsden for his very helpful feedback, and Dr. Lee Kelvin for his help with making RGB images. 





\bibliographystyle{mnras}
\bibliography{g316}

\appendix

\begin{table*}
\centering
\caption{Masers found in close proximity to G316.81--0.06.}
\label{tab:masers}
\begin{tabular}{llccr}
\hline
RA (J2000) & Dec (J2000) & Maser Type & v$_{\text{peak}}$ (km s$^{-1}$) & References \\
\hline 
14 45 26.6 & -59 49 14 & Hydroxyl & -36.7 & \citet{1967AuJPh..20..407M} \\ 
14 45 27.6 & -59 49 49 & Hydroxyl & -41 & \citet{1987AuJPh..40..215C} \\
14 45 26.34 & -59 49 15.4 & Hydroxyl & -44 & \citet{1998MNRAS.297..215C} \\
14 45 26.34 & -59 49 15.4 & Hydroxyl & -43.5 & \citet{2010MNRAS.406.1487B} \\
14 45 27.6 & -59 49 49 & Class II methanol & -44 & \citet{1992MNRAS.256..519M} \\ 
14 45 27.9 & -59 49 13 & Class II methanol & -42.1 & \citet{1995PASA...12...37C} \\
14 45 27.9 & -59 49 13 & Class II methanol & -46.8 & \citet{1995MNRAS.272...96C} \\
14 45 27.9 & -59 49 13 & Class II methanol & -45.7 & \citet{1995MNRAS.274.1126C} \\
14 45 28 & -59 49 12 & Class II methanol & -46.0 & \citet{1997MNRAS.291..261W} \\
14 45 26.44 & -59 49 16.3 & Class II methanol & -42.2 & \citet{1998MNRAS.301..640W} \\
14 45 26.44 & -59 49 16.5 & Class II methanol & -44.9 & \citet{1998MNRAS.301..640W} \\
14 45 26.44 & -59 49 16.4 & Class II methanol & -45.8 & \citet{1998MNRAS.301..640W} \\
14 45 26.44 & -59 49 16.3 & Class II methanol & -46.9 & \citet{1998MNRAS.301..640W} \\
14 45 26.44 & -59 49 16.3 & Class II methanol & -48.1 & \citet{1998MNRAS.301..640W} \\
14 45 26.4 & -59 49 16.5 & Class II methanol & -46.0 & \citet{2005AA...432..737P} \\
14 45 26.4 & -59 49 16.3 & Class II methanol & -46.3 & \citet{2009PASA...26..454C} \\
14 45 26.4 & -59 49 16.3 & Class II methanol & -46.3 & \citet{2010MNRAS.406.1487B} \\
14 45 26.4 & -59 49 16.3 & Class II methanol & -45.8 & \citet{2012MNRAS.420.3108G} \\ 
14 45 30.3 & -59 51 52 & Water & -48.6 & \citet{1976Natur.260..306K} \\
14 45 25.0 & -59 49 31 & Water & -46 & \citet{1980AuJPh..33..139B} \\ 
14 45 26.58 & -59 49 14.1 & Water & -46 & \citet{2010MNRAS.406.1487B} \\
14 45 25.5 & -59 49 18 & Water & -47.2 & \citet{2011MNRAS.416.1764W} \\
14 45 26.1 & -59 49 19.5 & Water & -50.6 & \citet{2014MNRAS.442.2240W} \\
14 45 26.4 & -59 49 15.3 & Water & -45.5 & \citet{2014MNRAS.442.2240W} \\
14 45 26.4 & -59 49 15.2 & Water & -40.9 & \citet{2014MNRAS.442.2240W} \\ 
14 45 26.4 & -59 49 15.5 & Water & -39.2 & \citet{2014MNRAS.442.2240W} \\ 
14 45 26.4 & -59 49 15.4 & Water & -34.6 & \citet{2014MNRAS.442.2240W} \\ 
\hline
\end{tabular}
\end{table*}

\bsp	
\label{lastpage}
\end{document}